\documentclass[11pt]{article}

\usepackage{amssymb,amsthm,amscd,array,stmaryrd,xypic,url}

\usepackage[T1]{fontenc}
\usepackage[english]{babel} 
\usepackage[latin1]{inputenc}

\let\ssection=\section
\renewcommand{\section}{\setcounter{equation}{0}\ssection}

\def\a{\alpha}
\def\b{\beta}
\def\g{\gamma}
\def\Ga{\Gamma}
\def\d{\delta}

\def\k{\kappa}
\def\l{\lambda}

\def\m{\mu}
\def\om{\omega}
\def\Om{\Omega}

\newcommand{\bbR}{\mathbb{R}}

\newcommand{\bbC}{\mathbb{C}}

\newcommand{\bbN}{\mathbb{N}}
\newcommand{\bbZ}{\mathbb{Z}}

\newcommand{\bbS}{\mathbb{S}}

\newcommand{\bi}{\mathsf{i}}

\newcommand{\Ber}{\mathrm{Ber}}

\newcommand{\cC}{{\mathcal{C}}}

\newcommand{\ce}{\mathrm{ce}}

\newcommand{\Cinfty}{{\mathcal{C}^{\infty}}}

\newcommand{\Cl}{\mathrm{Cl}}
\newcommand{\bbCl}{\mathrm{\bbC l}}

\newcommand{\conf}{\mathrm{conf}}

\newcommand{\sD}{\mathsf{D}}
\newcommand{\D}{\mathcal{D}}

\newcommand{\Dlm}{\mathcal{D^{\l,\m}}}
\newcommand{\Dslm}{\mathsf{D^{\l,\m}}}

\newcommand{\cE}{\mathcal{E}}

\newcommand{\e}{\mathrm{e}}
\newcommand{\ev}{\mathsf{ev}}

\newcommand{\End}{\mathrm{End}}
\newcommand{\cF}{{\mathcal{F}}}
\newcommand{\sF}{{\mathsf{F}}}
\newcommand{\fkg}{{\mathfrak{g}}}
\newcommand{\gl}{{\mathrm{gl}}}
\newcommand{\GL}{{\mathrm{GL}}}
\newcommand{\GQ}{{\mathrm{GQ}}}

\newcommand{\Gr}{{\mathrm{Gr}}}
\newcommand{\tg}{\tilde{\gamma}}

\newcommand{\calH}{\mathcal{H}}
\newcommand{\cJa}{\mathcal{J}^0}
\newcommand{\cJb}{\mathcal{J}^1}

\newcommand{\cI}{\mathcal{I}}
\newcommand{\Id}{\mathrm{Id}}

\newcommand{\ccL}{\mathcal{L}}
\newcommand{\sL}{\mathsf{L}}
\newcommand{\LD}{\mathcal{L}_X^{\lambda,\mu}}
\newcommand{\Ll}{\ell_X^\lambda}

\newcommand{\sLl}{\mathsf{L}_X^\lambda}
\newcommand{\sLm}{\mathsf{L}_X^\mu}
\newcommand{\Ld}{L_X^\delta}
\newcommand{\bbL}{\mathbb{L}}

\newcommand{\Lt}{\mathbb{L}_X^\delta}
\newcommand{\ro}{\mathrm{o}}

\newcommand{\Obs}{\mathrm{Obs}}

\newcommand{\btN}{\hat{N}}
\newcommand{\btP}{\hat{P}}
\newcommand{\cPN}{\mathcal{P}_N}
\newcommand{\Pol}{\mathrm{Pol}}
\newcommand{\pol}{\mathrm{pol}}
\newcommand{\PreQ}{{\mathrm{PreQ}}}
\newcommand{\calP}{{\mathcal{P}}}
\newcommand{\cQ}{{\mathcal{Q}}}
\newcommand{\cQlm}{\mathcal{Q}^{\l,\m}}

\newcommand{\cM}{{\mathcal{M}}}
\newcommand{\cN}{{\mathcal{N}}}
\newcommand{\rO}{\mathrm{O}}

\newcommand{\cS}{{\mathcal{S}}}

\newcommand{\rST}{\mathrm{S}_{\cT}}
\newcommand{\sS}{\mathsf{S}}

\newcommand{\spin}{\mathrm{spin}}

\newcommand{\Spin}{\mathrm{Spin}}

\newcommand{\cT}{\mathcal{T}}

\newcommand{\ve}{\varepsilon}
\newcommand{\Vect}{\mathrm{Vect}}
\newcommand{\vol}{\mathrm{vol}}

\newcommand{\bzeta}{\hat{\zeta}}

\newcommand{\tp}{\tilde{p}}
\newcommand{\txi}{\tilde{\xi}}

\newcommand{\half}{\frac{1}{2}}

\newcommand{\medbox}[1]{\fbox{%
\rule[-10pt]{0pt}{25pt}$\;\;\displaystyle{#1}\;\;$}%
}

\begin{document}

\baselineskip=15pt



\newtheorem{thm}{Theorem}[section]
\newtheorem{lem}[thm]{Lemma}
\newtheorem{cor}[thm]{Corollary}
\newtheorem{prop}[thm]{Proposition}
\newtheorem{defi}[thm]{Definition}
\newtheorem{ex}[thm]{Example}
\newtheorem{rmk}[thm]{Remark}

\title{Conformal geometry\\ of the\\ supercotangent and spinor bundles}

\author{
J.-P.~Michel\thanks{
\textsc{Institut Camille Jordan, Universit\'e Claude Bernard Lyon $1$, $43$ boulevard du $11$ novembre $1918$, F-$69622$ Villeurbanne Cedex France};
E-mail: jpmichel@math.univ-lyon1.fr; Tel.: +33 (0)4 72 44 79 41; Fax: +33 (0)4 72 43 16 87 .
}
}

\maketitle

\thispagestyle{empty}

\begin{abstract}
We study the actions of local conformal vector fields $X\in \conf(M,g)$ on the spinor bundle of $(M,g)$ and on its classical counterpart: the supercotangent bundle $\cM$ of $(M,g)$. We first deal with the classical framework and determine the Hamiltonian lift of $\conf(M,g)$ to $\cM$. We then perform the geometric quantization of the supercotangent bundle of $(M,g)$, which constructs the spinor bundle as the quantum representation space. The Kosmann Lie derivative of spinors is obtained by quantization of the comoment map.

The quantum and classical actions of $\conf(M,g)$ turn, respectively, the space of differential operators acting on spinor densities and the space of their symbols into $\conf(M,g)$-modules. They are filtered and admit a common associated graded module. In the conformally flat case, the latter helps us determine the conformal invariants of both $\conf(M,g)$-modules, in particular the conformally odd powers of the Dirac operator. 
\end{abstract}



\section{Introduction}

Conformal geometry naturally emerges in physics from the study of the dynamics of free massless particles in space-time $(M,g)$. This has for the quantum counterpart the conformal invariance of the wave operator and the Dirac operator, describing the dynamics of the free massless fields 
of spin $0$ and $\half$ respectively. 

On a pseudo-Riemannian manifold $(M,g)$, the conformal invariance of an object means its invariance under a rescaling $g\mapsto F g$ of the metric, where $F$ is a positive function on $M$. In the conformally flat case, this is equivalent to invariance under the action of $\conf(M,g)$, the sheaf of local conformal vector fields on $(M,g)$. The latter is then locally isomorphic to the conformal Lie algebra $\ro(p+1,q+1)$, with $(p,q)$ the signature of $g$. A great part of conformal geometry is precisely the study of conformal invariants, and those generalizing the wave operator have been intensively investigated.
 In the conformally flat case, Eastwood and Rice \cite{ERi87} have classified conformally invariant operators in a general setting, and obtained in particular those with values in tensor densities, e.g. the wave operator (or Yamabe operator). They are the conformal powers of the Laplacian and have been generalized later as the GJMS operators \cite{GJMS92} in the curved case. Since then, they have attracted much attention, especially their zeroth-order term which generates the celebrated $Q$-curvature \cite{Bra95}. Their classical counterparts are much simpler and consist in their principal symbols only, which happen to be the powers of the free Hamiltonian on $T^*M$. In the conformally flat case, this correspondence between quantum and classical conformal invariants can be enlarged to an isomorphism of $\ro(p+1,q+1)$-modules, between the space of differential operators acting on densities and the space of their symbols. This isomorphism is a symbol map whose inverse is the so-called conformally equivariant quantization \cite{DLO99}. 
Quite recently, this quantization procedure has been generalized to the curved case \cite{MRa09,Rad09,Sil09}, exhibiting tight links with the GJMS operators~\cite{CGS10} and their symmetries \cite{Eas05,ELe08,GSi09}. The latter are of first importance from the point of view of integrability. 

It seems natural to ask for a spin analog of the above picture, where the Dirac operator replaces the wave operator, with in particular the hope to obtain new conformal invariants. Such a program has been initiated in \cite{Mic09}, and the present work should constitute its cornerstone, for the conformally flat case. 
Considering a spinning particle on configuration space (or space-time) $(M,g)$, the aim of this paper is to determine the actions of $\conf(M,g)$ on the classical and quantum phase spaces, and, next, to study the $\conf(M,g)$-modules of corresponding observables, namely the module of differential operators acting on spinor densities and the module of their symbols. Further results, on conformally equivariant quantization as well as on its eventual links with conformally invariant operators and their symmetries, can be found in \cite{Mic09} and will be the subject of forthcoming papers.   

For a spinning particle on $(M,g)$, the quantum framework is well-known. The "phase space" is the space of spinor fields, and differential operators acting on them constitute a space of observables. As classical phase space we choose the supercotangent bundle of $M$, given by the fibered product $\cM=T^*M\times_M\Pi TM$ with $\Pi$ the reverse parity functor. In~the flat case, such a choice can be traced back to Berezin and Marinov \cite{BMa77}, and Casalbuoni et al. \cite{BCL77}. The geometric definition of $\cM$ was first given by Getzler \cite{Get83}, as the geometric realization of the algebra of symbols of the differential operators acting on spinor fields. The non-usual Grassmann component of the algebra of symbols comes from the dequantization of the Clifford algebra acting on spinors.  Rothstein gave, later, a representation theorem of the even symplectic structures on a supermanifold \cite{Rot90}, which yields a canonical symplectic form on the supercotangent bundle of $(M,g)$. This allows one to deal with Hamiltonian mechanics of spinning particles (see e.g. \cite{Mic09} for the recovering of the Papapetrou equations \cite{Pap51}), and complete the classical setting. To link the latter with the quantum one, geometric quantization is perfectly suited, since its extension to supermanifolds is well-established \cite{Kos77,Tuy92,ENi96}. As a first step, supermanifolds over one point have been quantized in that way by Voronov \cite{Vor90}: the resulting quantum representation space is the spinor module, whose usual construction is recovered in the framework of geometric quantization. This sounds quite promising, but only the prequantization of the supercotangent bundle has been performed yet~\cite{Rot90}. We bridge the gap in Section $4$ and prove that, as desired, geometric quantization of the supercotangent bundle leads to the spinor bundle as a quantum representation space. 

Hamiltonian actions of conformal vector fields on the supercotangent bundle have, up to our knowledge, never been even defined. On the contrary, there is a whole literature dealing with the action of conformal vector fields on spinor bundles, defining a Lie derivative of spinors. In her seminal work \cite{Kos72}, Kosmann provides a construction of such an action, natural from a geometric point of view, but by no means canonical, and several interpretations of it have been proposed, more geometrically \cite{BGa92,GMa05} or physically \cite{PWi08} rooted. Nevertheless, no other definition has been proposed for the Lie derivative of spinors along conformal vector fields. Taking advantage of its own definition, Kosmann has established in \cite{Kos72} the invariance of the Dirac operator under the action of~$\conf(M,g)$. This work has not been extended to higher-order operators, and the expected conformal invariance of some powers of the Dirac operator seems to have never been considered, until the work of Holland and Sparling, in terms of rescalings of the metric \cite{HSp01}. 

The main results of this paper are: the correspondence between the new Hamiltonian $\conf(M,g)$-action on $\cM$ and the Kosmann Lie derivative of spinors provided by geometric quantization of $\cM$, the comparison of the $\conf(M,g)$-module of spinor differential operators with its two modules of symbols and the identification of its graded Poisson algebra, and finally the classification of their conformal invariants in the conformally flat case. Let us detail the content of the present work.

We take advantage of Section $2$ to introduce the needed elements of spin geometry and supergeometry. 
We prove, in particular, that Clifford algebras arise as the Moyal-Weyl quantization of symplectic supermanifolds over one point.

Next, we introduce in Section $3$ the supercotangent bundle of $(M,g)$ together with its symplectic structure, given by an exact and even $2$-form $d\a$. In contradistinction with the case of the cotangent bundles, it proves to be non-trivial to lift $\conf(M,g)$ to $\cM$ in a Hamiltonian way. The natural requirement to preserve $\a$ does not provide an unique lift of $\Vect(M)$, so we further demand the preservation of an exact and odd symplectic form $d\b$ on $\Pi TM$. This enables us to get a unique Hamiltonian lift, but defined for a Lie subalgebra of $\Vect(M)$ only. The latter coincides with $\conf(M,g)$ for our choice of $d\beta$ as the pull-back to $\Pi TM$ of the canonical odd symplectic form of $\Pi T^*M$ via the metric $g$. Finally, we give the even and odd comoment maps $\cJa$ and $\cJb$ of this Hamiltonian action of $\conf(M,g)$ on $\cM$.  

We start Section $4$ with a brief reminder of geometric quantization. Our aim is to develop that of the supercotangent bundle $\cM$, which is essentially the merging of geometric quantization of the cotangent bundle $T^*M$ and that of a supermanifold over one point. We develop the latter in some extent, generalizing to arbitrary metrics the work of Voronov  \cite{Vor90}, stated for an Euclidean one. We complete then the geometric quantization of $\cM$ upon topological conditions on $M$, which amounts to restrict to almost-complex spin manifolds if the metric $g$ is Riemannian. We end up with the following theorem, which is a merging of Theorems \ref{thmSpinorB} and \ref{thmSpinorL} and provides a new interpretation for the Kosmann Lie derivative of spinors.
\begin{thm}
The geometric quantization of $(\cM,\om)$ constructs the spinor bundle $\sS$ of $(M,g)$, and $\Ga(\sS)$ is the quantum representation space. The geometric quantization mapping $\cQ_{\GQ}$ establishes the correspondence
$$
\medbox{\forall X\in\conf(M,g),\quad\cQ_{\GQ}(\cJa_X)=\frac{\hbar}{\bi}\sL_X,} 
$$
where $\sL_X$ is the Lie derivative of spinors along $X$, proposed by Kosmann \cite{Kos72}. 
\end{thm}

In Section $5$, we define the $\conf(M,g)$-module $\Dslm$ of differential operators acting on spinor densities, endowed with the adjoint action of the Lie derivative of spinors, and the $\conf(M,g)$-module $\cS^\d[\xi]$ of $\d$-weighted functions on $\cM$, endowed with the afore-mentioned Hamiltonian action on $\cM$. The both modules admit so-called natural and Hamiltonian filtrations. We prove that,  for the Hamiltonian filtration, $\cS^\d[\xi]$ is the graded module associated to $\Dslm$ and if $\l=\m$ it is also its graded Poisson algebra. We restrict then to conformally flat manifolds $(M,g)$ until the end of the section. This enables us to prove that $\Dslm$ and $\cS^\d[\xi]$ have a common associated graded module $\cT^\d[\xi]$ of tensorial symbols, for the natural filtration. That is quite different from the case of scalar differential operators, where both filtrations and spaces of symbols prove to be the same  (see e.g. \cite{DLO99}). Our aim turns then to the classification of the conformal invariants of the three preceding modules $\cT^\d[\xi]$, $\cS^\d[\xi]$ and $\Dslm$, for $M$ an oriented manifold. Adopting the strategy of \cite{ORe03}, we compute explicitly the action of generators of $\conf(M,g)$, and use Weyl's theory of invariants \cite{Wey97}. We end up with a full classification of the conformal invariants among spinor differential operators and their symbols, recovering in particular the following theorem, where $\g^i$ are the Clifford matrices, $\vol_g$ the volume form of $(M,g)$ and $(x^i,p_i,\xi^i)$ coordinates on $\cM$.
\begin{thm}
The conformally invariant differential operators in $\Dslm$ are, 
\begin{enumerate}
\item the chirality: $(\vol_g)_{i_1\cdots i_n}\g^{i_1}\ldots\g^{i_n}\in\sD^{\l,\l}$,
\item the Dirac operator $\g^i\nabla_i$
and its twist $g^{ij_1}(\vol_g)_{j_1\ldots j_n}\g^{j_2}\ldots \g^{j_n}\nabla_{i}$, in $\sD^{\frac{n-1}{2n},\frac{n+1}{2n}}$,
\item for $s\in\bbN^*$, the operator in $\sD^{\frac{n-2s-1}{2n},\frac{n+2s+1}{2n}}$ given locally by $\cN(\Delta\, R^s)$, with $R=g^{ij}p_ip_j$, $\Delta=p_i\xi^i$ and $\cN$ the normal ordering (see \ref{OrdreNormal}).
\end{enumerate}
\end{thm}

Section $6$ gives us the opportunity to present some open questions and to draw several perspectives, that will be investigated in future papers. 

We use the Einstein conventions and freely lower and rise indices of coordinates, vector fields and tensor fields thanks to the metric $g$ on $M$.

%

\section{Preliminaries}
We present in this section the basic definitions and notation used throughout this paper. The algebra of differential operators acting on spinors and its graded algebra of symbols are introduced, as well as supermanifolds, in particular over one point. 

\subsection{Elements of spin geometry}
We define here the Clifford algebra and its spin module, prior to geometrizing them as fibers of bundles over a manifold. This is a classical subject extensively treated in the literature, see e.g. \cite{LMi89,BGV92}.

\subsubsection{Algebraic structures}

Clifford algebras are algebras canonically associated to metric vector space $(V,g)$, where $g$ is a symmetric and non-degenerate bilinear form of given signature. They are defined by $\Cl(V,g)=\cT(V)/\cI(V,g)$, i.e$.$ the quotient 
of the tensor algebra of $V$ by the ideal generated by the Clifford relations $u\otimes v+v\otimes u+2g(u,v)$, for $u,v\in V$. 
The gradation of the tensor algebra induces a filtration on the Clifford algebra: $\Cl_0(V,g)\subset \Cl_1(V,g)\subset \cdots \subset \Cl_n(V,g)=\Cl(V,g)$, where $n$ is the dimension of $V$ and $\Cl_k(V,g)$ is the above quotient of the vector space of tensors of order $k$ at most. As any filtered algebra, it admits an associated graded algebra, which is by definition $\Gr\,\Cl(V,g)=\bigoplus_{k=0}^{n}\Cl_k(V,g)/\Cl_{k-1}(V,g)$, and that proves to be isomorphic to the Grassmann algebra over $V$,
\begin{equation}\label{GrCl}
\Gr\, \Cl(V,g)\simeq \Lambda V.
\end{equation}
Besides, there exists a unique irreducible space of representation of the Clifford algebra, called the spin module and denoted by $S$. For the complexified  Clifford algebra $\bbCl(V,g):=\Cl(V,g)\otimes\bbC$ and $V$ of even dimension, it satisfies $\bbCl(V,g)\simeq\End(S)$. 

\subsubsection{Geometric structures}\label{Sec212}

The preceding algebraic constructions can be geometrized over a pseudo-Riemannian manifold~$(M,g)$, supposed here of even dimension for simplicity. The Clifford bundle $\Cl(M,g)$ always exists and is unique, as an associated  bundle to the one of orthonormal frames. Its fiber in $x\in M$ is the Clifford algebra of $T^*_xM$ endowed with the metric $g_x^{-1}$ at $x\in M$. The algebra $\Ga(\Cl(M,g))$ of sections of the Clifford bundle is filtered and the geometric version of (\ref{GrCl}) reads 
$
\Gr\, \Ga(\Cl(M,g))\simeq \Om(M),
$
 where $\Om(M)$ is the space of differential forms over $M$. On the contrary, the spin bundle does not always exist. It is defined as the vector bundle $\sS\rightarrow M$ satisfying, 
\begin{equation}\label{Spin}
\End(\sS)\simeq\bbCl(M,g).
\end{equation} 
The spin bundle can be obtained as an associated bundle to a spinor frame bundle, with structural group $\Spin$ or $\Spin^c$ for example \cite{Tra08}. We suppose from now on that $M$ admits a spin bundle~$\sS$.

Given a vector bundle, we can define the algebra of differential operators acting on the sections of that bundle. Starting with the spin bundle $\sS$, we obtain $\D(M,\sS)$ the space of differential operators acting on spinors, i.e$.$, on sections of $\sS$. This algebra is filtered by the order of derivations over the subalgebra $\Ga(\End \sS)\cong\Ga(\bbCl(M,g))$ of zeroth order operators. Since the latter admits also a filtration, we get a bifiltration of $\D(M,\sS)$ by the following subspaces, indexed by $k\in\bbN$ and $\k\leq n$,
$$
\D_{k,\k}(M,\sS)=\text{span}\{A(x)\nabla_{X_1}\cdots\nabla_{X_m}\,|\;m\leq k, A\in\Ga(\bbCl_{\k'}(M,g)) \text{ with } \k'=\k+2(k-m)\},
$$
where $\nabla_{X_i}$ is a spinor covariant derivative along the vector field $X_i$. The somehow strange condition on $\k'$ makes the latter space independent of the chosen connection. 
Generically, the graded algebra associated to an algebra of differential operators is called an algebra of symbols, that of scalar differential operators $\D(M)$ is $\Gr\, \D(M)\simeq \Ga(\cS TM)\simeq \Pol(T^*M)$, where $\Ga(\cS TM)$ are the symmetric contravariant tensor fields over $M$ and $\Pol(T^*M)$ the algebra of functions on $T^*M$ which are fiberwise polynomial. We denote this algebra of symbols by $\cS(M)$. Usually, only the filtration of $\D(M,\sS)$ w.r.t.\ the order is considered, and its algebra of symbols is then $\cS(M)\otimes\Ga(\bbCl(M,g))$. Following Getzler \cite{Get83}, we go one step further and take rather into account its above bifiltration, leading to
\begin{equation}\label{SMxi}
\mathrm{Bigr}\,\D(M,\sS)\simeq \cS(M)\otimes\Om_\bbC(M),
\end{equation}  
as the algebra of symbols of $\D(M,\sS)$.
The latter can be interpreted as an algebra of tensors, or as an algebra of functions on the supercotangent bundle of $M$, see below.

\subsection{Basics of supergeometry}
 Supergeometry relies on the notion of supercommutative algebras, which are associative algebras endowed with a $\bbZ_2$-gradation, denoted by $|\cdot|$, such that
$
 ab=(-1)^{|a||b|}ba,
$
 for homogeneous elements $a$ and $b$ of the algebra. The elements of gradation $0$ are called even and the ones of gradation $1$ are said to be odd.
 
There are mainly two approaches to supermanifolds, \cite{Lei80,Kos77} and \cite{DeW92,Tuy04}. We use the first one in terms of sheaves, but in a very concrete setting relying on vector bundles $E\rightarrow M$ and the reverse parity functor $\Pi$. We only work with supermanifolds defined as  
$
\Pi E=(M,\Ga(\cdot,\Lambda E^*)) 
$, 
where $\Ga(\cdot,\Lambda E^*)$ is the sheaf of sections of the exterior bundle. The algebra of smooth functions on $\Pi E$ is $\Cinfty(\Pi E)=\Ga(M,\Lambda E^*)$, and it admits local coordinates $(x^i,\xi^a)$, where $(x^i)$ are even and form a coordinate system on $M$ and $(\xi^a)$ are odd and form a base of the fibers of $E^*$.

All the usual objects of differential geometry can be generalized in the framework of supermanifolds, in particular differential forms. They constitute a sheaf of bigraded algebras, by the cohomological degree $p(\cdot)$ and by the $\bbZ_2$-gradation $|\cdot|$, with the following law of commutation for homogeneous elements:
\begin{equation}\label{CommLaw}
\a\wedge\b=(-1)^{p(\a)p(\b)+|\a||\b|}\b\wedge\a.
\end{equation} 
As a consequence, for odd functions $\xi^a$ and $\xi^b$, we have $d\xi^a\wedge d\xi^b=d\xi^b\wedge d\xi^a$. 
 
 A symplectic structure on a supermanifold is a closed and non-degenerate $2$-form, which is even if not stated otherwise. Kostant has proved an extension of the Darboux theorem in that setting.
 \begin{thm}\cite{Kos77}\label{Darboux}
 Let $(\cM,\om)$ be a symplectic supermanifold. At any point, there exists a flat metric given by $\eta_{ab}=\pm\d_{ab}$, and local coordinates $(\tilde{x}^i,\tp_i,\txi^a)$, with $\tilde{x}^i,\tp_i$ even and $\txi^a$ odd, such that
\begin{equation}\label{omDarboux}
 \om=d\tp_i\wedge d\tilde{x}^i+\eta_{ab}d\txi^a\wedge d\txi^b.
\end{equation}
 \end{thm}
Such coordinates $(\tilde{x}^i,\tp_i,\txi^a)$ are called Darboux coordinates. The local model for a symplectic supermanifold is then $T^*\bbR^n\times\Pi \bbR^p$, where $T^*\bbR^n$ is endowed with its canonical symplectic structure and $\Pi\bbR^p$ is endowed with the symplectic structure associated to a flat metric on $\bbR^p$ of given signature. 
 
Let us mention the existence of odd symplectic structures, whose local model is the cotangent bundle with reverse parity, namely $(\Pi T^*M,d\xi_i\wedge dx^i)$ with $(x^i,\xi_i)$ a coordinate system of $\Pi T^*M$ \cite{Lei77}. 
 
\subsection{Symplectic supermanifolds over one point}\label{SecPiV}
Prior to the study of the supercotangent bundle, let us focus attention on the purely odd supermanifolds, which are essentially Grassmann algebras. We investigate their symplectic structures and the deformation of their algebras of functions by the Moyal product, which leads to Clifford algebras. This is well-known, and develop for example in \cite{FLl00,MPU08}.

A symplectic supermanifold over one point is a real metric vector space with reverse parity, $\Pi V$, whose algebra of functions is the Grassmann algebra $\Lambda V^*$. 
A system of coordinates on $\Pi V$ is provided by a dual basis $(\xi^1,\ldots,\xi^n)$ of $V$, $n$ being the dimension of $V$. It generates the superalgebra of functions of $\Pi V$, whose $\bbZ_2$-grading comes from the natural $\bbZ$-grading of~$\Lambda V^*$.

From the commutation law (\ref{CommLaw}) of differential forms, we readily deduce that a symplectic form $\om$ on $\Pi V$ is given by $\om=g_{ij}d\xi^i\wedge d\xi^j$ for $g$ a metric on $V$.
Darboux coordinates for such a symplectic form $\om$ are given by an orthonormal cobasis of $(V,g)$, and the signature of the metric is then a symplectic invariant. For quantization purposes (see the proposition below), we introduce a factor $\frac{\hbar}{2\bi}$ in the symplectic form of $\Pi V$, i.e$.$ $\om=\frac{\hbar}{2\bi}g_{ij}d\xi^i\wedge d\xi^j$. This leads to the Poisson bracket
$
\{\xi^i,\xi^j\}=-\frac{\bi}{\hbar}g^{ij},
$
 which is given by the bivector $\pi=\frac{\bi}{2\hbar}g^{ij}\partial_{\xi^i}\otimes\partial_{\xi^j}$. Following \cite{MPU08}, we introduce $m_t^\star=m_\wedge\circ\,\exp(t\frac{\hbar}{2\bi}\pi)$, the deformation of the exterior product $m_\wedge$ on $\Lambda V^*$ in the direction of $\pi$. As $\Lambda V^*$ is finite dimensional, $m^\star_1$ is well-defined, the exponential reducing to a finite sum. By analogy with $T^*\bbR^n$, the product $m_1^\star$ is denoted by $\star$ and called the Moyal product.  The following proposition gives a synthetic formulation of the properties of this star-product.
\begin{prop}\cite{Vor90,FLl00,MPU08}\label{gLie}
Let $(\Pi V,\om)$ be the symplectic supermanifold associated to the metric vector space $(V,g)$. The canonical embedding $\g:V^*\hookrightarrow\Cl(V^*,g^{-1})$, extends via the Moyal product $\star$ to an isomorphism of filtered algebra, equivariant w.r.t. $\rO(V^*,g^{-1})$,
\begin{eqnarray} \label{gLvCLv}
\g:(\Lambda V^*,\star)& \longrightarrow &\Cl(V^*,g^{-1})\\ \nonumber
\xi^{i_1}\star\cdots \star \xi^{i_\k}&\longmapsto &\frac{\g^{i_1}}{\sqrt{2}}\cdots\frac{\g^{i_\k}}{\sqrt{2}},
\end{eqnarray} 
where $\g^i=\g(\xi^i)$. 
For all $u\in\Lambda^0V^*\oplus\Lambda^1V^*\oplus\Lambda^2V^*=E$ and $v\in\Lambda V^*$, this isomorphism satisfies 
\begin{equation}\label{QWIsoLie}
\frac{\hbar}{\bi}\g(\{u,v\})=[\g(u),\g(v)].
\end{equation}
\end{prop}
\begin{proof}
Starting from the relation $\xi^i\star\xi^j+\xi^j\star\xi^i=\frac{\hbar}{\bi}\{\xi^i,\xi^j\}=-g^{ij}$, the universal property of Clifford algebras shows that the linear embedding $V^*\rightarrow(\Lambda V^*,\star)$, defined by $\xi^i\mapsto\sqrt{2}\xi^i$, extends to a unique algebra morphism $\Cl(V^*,g^{-1})\rightarrow(\Lambda V^*,\star)$. This is an isomorphism as it establishes a correspondence between generators and relations of these two algebras.
We define the extension of $\g$ as the inverse of this isomorphism, (\ref{gLvCLv}) follows. The Moyal product satisfying $\Lambda^kV\star\Lambda^lV\subset\oplus_{j=0}^{k+l}\Lambda^jV$, the map $\g$ preserves the filtration, and, since $\rO(V^*,g^{-1})$ acts by linear symplectomorphisms on $\Pi V$, this is also a morphism of $\rO(V^*,g^{-1})$-modules.

Let now $u\in E$ and $v\in\Lambda V^*$. By definition of $\g$, we obtain $[\g(u),\g(v)]=\g(u\star v-v\star u)$, and by definition of the Moyal product $u\star v-v\star u=\frac{\hbar}{\bi}\{u,v\}+\left(\frac{\hbar}{2\bi}\right)^2(\pi^2(u,v)-\pi^2(v,u))$. As $\pi^2$ is symmetric on $E$, this proves Equation (\ref{QWIsoLie}).
\end{proof}
This proposition makes precise the common assertion that Clifford algebras are deformation of Grassmann algebras. As in the well-known even case of $\cS(\bbR^n)=\Pol(T^*\bbR^n)$, the deformed algebra is a filtered algebra whose associated graded algebra is the original algebra.

Thanks to the preservation of the filtration by $\g$, we deduce that the graded algebra $\Gr\, \Cl(V,g)$ is isomorphic to the Grassmann algebra $\Lambda V$. Moreover, as $\g$ is an isomorphism of $\rO(V^*,g^{-1})$-modules, we may extend it to the geometric framework, and recover the following well-known corollary. 
\begin{cor}\cite{LMi89,BGV92}\label{GradCliff}
Let $(M,g)$ be a pseudo-Riemannian manifold. There is an isomorphism of filtered algebras,
$
\g:\Om(M)\rightarrow\Ga(\Cl(M,g)),
$
which coincides with (\ref{gLvCLv}) at any point of $M$ and is called the Weyl quantization of $\Om(M)$.
\end{cor}

\section{Geometry of the supercotangent bundle}
This section is devoted to the study of the supercotangent bundle $\cM$ of a pseudo-Riemannian manifold $(M,g)$, together with its canonical symplectic structure. We investigate the existence and uniqueness of a Hamiltonian lift from $M$ to its supercotangent bundle, especially  of conformal vector fields. This enables us to define a comoment map on $\cM$. 
In all this section, the dimension of the manifold $M$ is denoted by $n$.
\subsection{The symplectic structure of the supercotangent bundle}\label{SubsectionSpcM}

\subsubsection{Differential aspects of the supercotangent bundle}
Let us recall that the graded algebra of symbols of spinor differential operators is $\Gr \, \D(M,\sS)=\cS(M)\otimes\Om_\bbC(M)$. In order to interpret these symbols as functions, Getzler introduced the supercotangent bundle.
\begin{defi} \cite{Get83}
The supercotangent bundle of a manifold $M$ is the fibered product {$\cM=T^*M\times_M \Pi TM$}. Its base manifold is the cotangent bundle $T^*M$, and its superalgebra of functions is $\cC^\infty(\cM)=\cC^\infty(T^*M)\otimes\Om(M)$. 
\end{defi}
\begin{rmk}\label{CoordNatcM}
Starting from a (local) coordinate system $(x^i)$ on $M$, we can construct a natural coordinate system $(x^i,p_i,\xi^i)$ on $\cM$, where $p_i$ and $\xi^i$ correspond respectively to $\partial_i$ and $dx^i$.
\end{rmk}
 
The graded algebra of symbols of $\D(M,\sS)$, introduced in (\ref{SMxi}), can now be interpreted as the algebra 
$
\cS(M)[\xi]:=\cS(M)\otimes\Om_\bbC(M),
$ 
of complex functions on $\cM$ which are polynomial in the fiber variables. 
It admits a bigradation given by the degree in the even and odd fiber variables, which is expressed via the decomposition 
$
\cS(M)[\xi]=\bigoplus_{k=0}^\infty\bigoplus_{\k=0}^n\cS_{k,\k}(M)[\xi]
$. 
The subspace $\cS_{k,\k}(M)[\xi]$ is isomorphic to the space of tensors $\Ga(\cS^kTM\otimes\Lambda^\k T^*M)$ via the canonical map
\begin{eqnarray}\label{Tenseur_Poly}
\Ga\left(\cS^k TM \otimes\Lambda^\k T^*M\right)&\rightarrow& \cS_{k,\k}(M)[\xi]\\ \nonumber
P^{i_1 \dots i_k}_{j_1 \ldots j_\k}(x)\,  dx^{j_1}\wedge\ldots \wedge dx^{j_\k}\,\otimes\,\partial_{i_1}\odot\ldots \odot\partial_{i_k} &\mapsto&
P^{i_1 \dots i_k}_{j_1 \ldots j_\k}(x)\,  \xi^{j_1}\ldots \xi^{j_\k}\,p_{i_1}\ldots p_{i_k}. 
\end{eqnarray}

Now, we define covariant derivatives and $1$-forms on $\cM$. Starting with a natural coordinate system $(x^i,p_i,\xi^i)$ on $\cM$, we denote by  $(\partial_i,\partial_{p_i},\partial_{\xi^i})$ the associated basis of local (left) derivations of $\cC^\infty(\cM)$ and by $(dx^i,dp_i,d\xi^i)$ its dual basis. 
Let $\Vect_V(\cM)$ be the space of vertical derivations of $\cM$ over $M$, locally generated by $(\partial_{p_1},\ldots,\partial_{p_n},\partial_{\xi^1},\ldots,\partial_{\xi^n})$. We have the short exact sequence of left $\Cinfty(\cM)$-modules
$$
0\rightarrow\Vect_V(\cM)\rightarrow\Vect(\cM)\rightarrow \cC^\infty(\cM)\otimes \Vect(M)\rightarrow 0.
$$
Thanks to the Levi-Civita connection we can trivialize this exact sequence and define a canonical lift of the vector fields of $M$. Thus, we associate to a natural coordinate system a basis of derivations, transforming tensorially over $M$,
\begin{equation}\label{BaseDerivationTensorielle}
\partial^\nabla_i=\partial_i+\Gamma^k_{ij}p_k\partial_{p_j}-\Gamma^k_{ij}\xi^j\partial_{\xi^k},  \qquad \partial_{p_i} \quad \text{and}\quad \partial_{\xi^i},
\end{equation}
where $\Ga^k_{ij}$ denote the Christoffel symbols and $i,j,k=1,\ldots,n$. The dual basis, of the right $\Cinfty(\cM)$-module of $1$-forms on $\cM$, transforms also tensorially and is given by, 
\begin{equation}\label{DHxXiDual}
dx^i,\qquad d^\nabla p_i=dp_i-\Gamma^k_{ij}p_k dx^j \quad \text{and} \quad d^\nabla \xi^i=d\xi^i+\Gamma^i_{jk}\xi^j dx^k.
\end{equation}

\subsubsection{Symplectic aspects of the supercotangent bundle}

Let us recall that a symplectic form is an even closed $2$-form which is non-degenerate. By definition, $\Cinfty(\cM)$ admits a $\bbZ$-gradation in the odd variables, and a form will be called quadratic if its degree in odd variables is at most quadratic.
The following theorem is a particular case of the general results of Rothstein on symplectic supermanifolds, rediscovered by Bordemann \cite{Bor00} from a deformation quantization point of view and by Roytenberg \cite{Roy02}.
\begin{thm}\cite{Rot90}
Let $\cM$ be the supercotangent bundle of a manifold $M$. There is a $1$-$1$ correspondence between
\begin{enumerate}
 \item Symplectic forms $\om$ of quadratic type on $\cM$, 
\item Pseudo-Riemannian metrics $g$ on $M$, together with a $g$-compatible connection $\nabla$ on $TM$.
\end{enumerate}
\end{thm}
Moreover, Rothstein proves that every symplectic form on $\cM$ can be pulled-back to a quadratic one, via a diffeomorphism of $\cM$ preserving $x^i$ and $p_i$ and transforming $\xi^i$ in $\xi^i$ plus higher order terms in the odd coordinates. 
On the supercotangent bundle of $(M,g)$, the latter theorem yields to a canonical even symplectic form depending on the Levi-Civita connection as well as on its curvature, given by the Riemann tensor 
$R^c_{aij}=(\partial_i\Ga^c_{aj}-\partial_j\Ga^c_{ai})+(\Ga_{ai}^k\Ga_{jk}^c-\Ga_{aj}^k\Ga_{ik}^c)$.
\begin{cor}\cite{Rot90}
Let $(M,g)$ be a pseudo-Riemannian manifold, and $\nabla$ be its Levi-Civita connection. There is a canonical symplectic $2$-form on $\cM$, which is written in natural coordinates:
\begin{equation}\label{dalpha}
\medbox{\om=dp_i\wedge dx^i+\frac{\hbar}{4\bi}\, g_{lm}R^m_{kij}\xi^k\xi^l dx^i\wedge dx^j+\frac{\hbar}{2\bi}\, g_{ij}d^\nabla\xi^i\wedge d^\nabla \xi^j.}
\end{equation}
This is the exact differential of the potential $1$-form 
\begin{equation}\label{alpha}
\medbox{\a=p_idx^i+\frac{\hbar}{2\bi} g_{ij}\xi^i d^\nabla\xi^j.}
\end{equation}
\end{cor}
 The imaginary factor $\frac{\hbar}{2\bi}$ is introduced for quantization purposes. Since the symplectic form $\om$ is real, with the dimension of an action, the $\xi$ variables carry the same dimension as the Clifford matrices, and the conjugation is given by $\overline{\xi^i\xi^j}=\bar{\xi^j}\bar{\xi^i}=-\xi^i\xi^j$, in accordance with adjunction on Clifford matrices.
\begin{rmk}
The symplectic manifold $(\cM,\om)$ turns out to be the natural phase space to deal with classical spinning particles with configuration space $(M,g)$. With this standpoint, we can recover the Papapetrou equations for a spinning particle on $(M,g)$  \cite{Pap51}, as the equations of motion associated to the free Hamiltonian $g^{ij}p_ip_j$ on $(\cM,\om)$ \cite{Mic09}.  
\end{rmk}
If there is a metric $g$ on $M$, the odd symplectic form of $\Pi T^*M$ can be pulled-back to the supermanifold $\Pi TM$, and writes then as $d\b=g_{ij}d^\nabla\xi^i\wedge dx^j$, where $\b=g_{ij}\xi^idx^j$. 

\subsubsection{Darboux coordinates on $(\cM,\om)$}

We will determine explicit Darboux coordinates on $(\cM,\om)$, in terms of natural coordinates~$(x^i,p_i,\xi^i)$.  For this purpose, we introduce a local orthonormal frame field $(e_a)_{a=1,\ldots,n}$ on~$(M,g)$. It is obtained from the natural frame $(\partial_i)_{i=1,\ldots,n}$ by the change of frames $(e^i_a)$. Denoting its inverse by $(\theta^a_i)$, the Levi-Civita connection $1$-form takes the expression
\begin{equation}\label{om_ab}
\om_b^a=\theta^a_i\left(de^i_b+\Ga^i_{jk}e^j_bdx^k\right),
\end{equation}
where the $\Ga^i_{jk}$ are the Christoffel symbols of the Levi-Civita connection. 
\begin{prop}\label{PropCoordDarboux}
Let $(e_a)_{a=1,\ldots,n}$ be an orthonormal frame of $(M,g)$, and $(x^i,p_i,\xi^i)$ be natural coordinates of $\cM$. Darboux coordinates are given by the functions 
\begin{equation}\label{CoordDarboux}
\medbox{x^i,\qquad \txi^a=\theta^a_i\xi^i,\qquad \tp_i=p_i+\frac{\hbar}{2\bi}\om_{bi}^a\txi_a\txi^b,}
\end{equation}
where $\txi_a=e_a^ig_{ij}\xi^j$ and $\om_{bi}^a=\langle\partial_i,\om_b^a\rangle$. 
\end{prop}
\begin{proof}
Starting with definition (\ref{alpha}) of $\a$, we replace $d^\nabla\xi^j$ by its expression (\ref{DHxXiDual}) and use $\xi^j=e^j_b\txi^b$ in order to obtain,
$$
\a=p_idx^i+\frac{\hbar}{2\bi}(\Ga_{ik}^j\xi_j\xi^kdx^i+\theta^a_j\txi_a\txi^bde^j_b)+\frac{\hbar}{2\bi}\eta_{ab}\txi^ad\txi^b,
$$
where $\eta$ is the flat metric such that $\eta_{ab}=g_{ij}e^i_ae^j_b$. Taking advantage of the formula (\ref{om_ab}) for the Levi-Civita connection $1$-form, we end up with the expression $\a=\tp_idx^i+\frac{\hbar}{2\bi}\eta_{ab}\txi^ad\txi^b$, which gives after differentiation the required Darboux canonical form (\ref{omDarboux}) for $\om$.

The functions $(x^i,\tp_i,\txi^a)$ clearly generate $\Cinfty(\cM)$, and their associated derivations commute, since they are given by the Hamiltonian vector fields of $\tp_i$, $-x^i$ and $-\txi^a$. Hence, they form indeed a coordinate system of $\cM$.
\end{proof}

\subsection{Hamiltonian actions on the supercotangent bundle}\label{SubSecconfcM}
On the cotangent bundle $T^*M$, there is a unique Hamiltonian lift of every vector field ${X\in\Vect(M)}$. Moreover, this lift coincides with the natural lift of $X$ to $T^*M$. The construction of such a unique Hamiltonian lift of $X$ to the supercotangent bundle $\cM$ of $(M,g)$ is more problematic, and has not been considered (as far as we know) in the literature. The method presented here has been developed by Duval in the flat case. First, we compute the lifts $\hat{X}$ of $X\in\Vect(M)$ preserving the potential $1$-form $\a$ of $\cM$, given by (\ref{alpha}). We impose then an additional condition on $\hat{X}$, namely to preserve the direction of $\b$, the $1$-form defining the odd symplectic structure of $\Pi TM$. This allows us to insure the uniqueness of the lift, but, in return, only do the conformal vector fields admit such a lift. As a consequence we get an Hamiltonian action on $(\cM,\om)$ and compute the associated comoment map.

\subsubsection{Hamiltonian lift of conformal vector fields on $(M,g)$ to $\cM$}

\begin{lem}\label{LXaLem}
Let $(\cM,d\a)$ be the supercotangent bundle of $(M,g)$ endowed with its canonical symplectic structure. A lift $\hat{X}$ of the vector field $X\in\Vect(M)$ to the supercotangent bundle $\cM$, which preserves $\a$, retains the form
\begin{equation}\label{lemXY}
\hat{X}= X^i\partial^\nabla_i+Y_{ij}\xi^j\partial_{\xi_i}-p_j\nabla_i X^j\partial_{p_i}
+ \frac{\hbar}{2\bi} \left(R^k_{lij}\xi_k\xi^l  X^j- (\nabla_iY_{kl})\xi^k\xi^l \right)\partial_{p_i},
\end{equation}
where $Y$ is an arbitrary $2$-form on $M$, depending linearly on $X$.
\end{lem}
\begin{proof}
We look for the lifts $\hat{X}$ of $X$ which preserve $\a$, i.e$.$ vector fields of the form
$
\hat{X}=X^j\partial^\nabla_j+P_j\partial_{p_j}+\Xi^j\partial_{\xi^j},
$
and such that 
$
L_{\hat{X}} \alpha=0
$. 
 The Cartan formula still holds on supermanifolds and leads to
$d\langle \hat{X}, \alpha\rangle + \langle \hat{X}, d\alpha\rangle=0$. We use the fact that $dp_i\wedge dx^i=d^\nabla p_i\wedge dx^i$ and write the action of $d$ in covariant terms to obtain
\begin{eqnarray} \nonumber
0&=& \Big[P_i+p_j\nabla_i X^j+ \frac{\hbar}{2\bi} \left( \partial^\nabla_i(\xi^k\Xi_k)  
- R^k_{lij}\xi_k\xi^l  X^j)\right) \Big] dx^i\\ \label{LXa}
&&
+\Big[X^i+\frac{\hbar}{2\bi} g_{kl}\xi^k\partial_{p_i}\Xi^l-X^i\Big]d^\nabla p_i\\ \nonumber
&&
+\frac{\hbar}{2\bi} \Big[-g_{kl}\Xi^l+g_{lm}\xi^l\partial_{\xi^k}\Xi^m+2g_{kl} \Xi^l \Big]d^\nabla\xi^k.
\end{eqnarray}
As $(dx^i,d^\nabla p_i,d^\nabla\xi^i)$ are free over $\Cinfty(\cM)$, each of these three terms must vanish. The second one yields the independence of the $\Xi^i$ on the $p_j$, and the last one gives, for all ${i=1,\ldots,n}$,
\begin{equation}\label{Xi}
\Xi_i+\xi_j\partial_{\xi^i}\Xi^j=0.
\end{equation}
As $\Xi_i$ is an odd function, we may decompose it as: $\Xi_i=Y_{ij}\xi^j+Z_{ijkl}\xi^j\xi^k\xi^l+\ldots$, where $Y$ and $Z$ are even covariant tensor fields on $M$. By substitution in the Equation (\ref{Xi}), we get
$$
Y_{ij}\xi^j+Y_{ji}\xi^j+(Z_{ijkl}+3Z_{jikl})\xi^j\xi^k\xi^l+\cdots=0.
$$
The antisymmetry of $Y$ and the equation $Z_{ijkl}+3Z_{jikl}=0$ follow. To find $Z$, we antisymmetrize in $i$ and $j$ the last equality, and we find $2Z_{[ji]kl}=0$, i.e$.$ $Z_{ijkl}=Z_{jikl}$. Coming back to the initial equation, we conclude that $Z_{ijkl}=0$. The same could be applied to each tensor of higher order, leading thus to $\Xi_i=Y_{ij}\xi^j$ with $Y$ an even skew-symmetric $2$-tensor on $M$.
From the vanishing of the $dx^i$ factor in the equation (\ref{LXa}), we deduce
$$
-P_i=p_j\nabla_i X^j+ \frac{\hbar}{2\bi} \left( \partial^\nabla_i(Y_{jk}\xi^j\xi^k) - R^k_{lij}\xi_k\xi^l  X^j\right).
$$
By the definition (\ref{BaseDerivationTensorielle}) of the covariant derivative, we clearly have $\partial^\nabla_i(Y_{jk}\xi^j\xi^k)=(\nabla_iY_{kl})\xi^k\xi^l$. Together with the previously obtained expression of $\Xi_i$, we end up with the desired formula for~$\hat{X}$.
\end{proof}
The lift defined by Lemma \ref{LXaLem} is henceforth not unique, and the conditions that $Y$ has to satisfy in order to get a Lie algebra morphism are non trivial. We will overcome this difficulty by adding a further condition on the lift; namely that it preserves the direction of $\b=g_{ij}\xi^i dx^j$, which is the lift to $\cM$ of the canonical $1$-form of $\Pi TM$. As a drawback, we cannot lift every vector field in that way, but only those preserving the conformal structure of $(M,g)$. More precisely, we lift the sheaf $\conf(M,g)$ of local conformal vector fields $X$ on~$(M,g)$, defined by $L_Xg_{ij}=\l g_{ij}$ for some smooth function $\l$, depending on $X$. This is a sheaf of Lie algebras, and as we are always working locally, we consider $ \conf(M,g)$ as a Lie subalgebra of $\Vect(M)$. Remark that, on any supermanifold, the simultaneous preservation of odd and even symplectic forms yields to a finite dimensional subalgebra of vector fields \cite{Khu91}.
\begin{thm}\label{thmpb}
The conformal vector fields $X\in\conf(M,g)$ admit a unique lift to $\cM$ preserving~$\a$ and the direction of $\b$, given by
\begin{equation}\label{tildeXg}
\medbox{\tilde{X}=X^i\partial^\nabla_i+\partial_{[j} X_{i]}\xi^j\partial_{\xi_i}-p_j\nabla_i X^j\partial_{p_i}+ \frac{\hbar}{2\bi} \left(R^k_{lij}\xi_k\xi^l  X^j- (\nabla_i\partial_{[l} X_{k]})\xi^k\xi^l \right)\partial_{p_i},}
\end{equation}
where $\partial_{\xi_i}=g^{ij}\partial_{\xi_j}$ and the brackets denote skew-symmetrization. The mapping $X\mapsto \tilde{X}$ defines a Lie algebra morphism $\conf(M,g)\rightarrow \Vect(\cM)$.
\end{thm}

\begin{proof}
Let $X\in\Vect(M)$. Thanks to Lemma \ref{LXaLem}, we know that $\tilde{X}$ is of the form (\ref{lemXY}).
The undetermined tensor $Y$ will be fixed by preserving the direction of $\b$, i.e$.$  by the equation $L_{\tilde{X}}\beta=f\beta$, for some $f\in\Cinfty(M)$. Using the local expression $\beta=g_{ij}\xi^j dx^i$, we are led to
\begin{equation}\label{eq_LXb}
X(g_{ij})-g_{ik}X^l\Gamma^k_{jl}+Y_{ij}+g_{jk}\partial_i X^k=f g_{ij}.
\end{equation}
From the symmetrization of this equation in $i$ and $j$, it follows that $L_X g=2f g$. Thus, $X\in\conf(M,g)$, a condition we will assume from now on. For further reference, remark that 
\begin{equation}\label{LXBeta}
L_{\tilde{X}}\beta=\half\frac{L_Xg}{g}\beta.
\end{equation} 

We can as well antisymmetrize (\ref{eq_LXb}) in $i$ and $j$, which entails 
$
Y_{ij}=\partial_{[j} X_{i]}.
$
By substitution in formula (\ref{lemXY}), the expression (\ref{tildeXg}) of $\tilde{X}$ readily follows.
Besides, the conditions characterizing the lift ensure that $X\mapsto\tilde{X}$ is a Lie algebra morphism.
\end{proof}
In terms of Darboux coordinates $(x^i,\tp_i,\txi^i)$, introduced in Proposition \ref{PropCoordDarboux}, the $1$-forms $\a$ and $\b$ have the same expressions as in the flat case. As to the Hamiltonian lift $\tilde{X}$, its expression simplifies somehow, viz.,
\begin{equation}\label{tildeX}
\tilde{X}=X^i \tilde{\partial_i}+\half\left((\partial_j X^i)\txi^j\partial_{\txi^i}-(\partial_i X^j)
\txi_j\partial_{\txi^i}\right)-\tp_j\partial_i X^j\partial_{\tp_i}- \frac{\hbar}{2\bi} \txi_j\txi^k(\partial_i\partial_k X^j)\partial_{\tp_i},
\end{equation}
where $(\tilde{\partial}_i,\partial_{\tp_i},\partial_{\txi^i})$  denote the derivations associated to the coordinates $(x^i,\tp_i,\txi^i)$.

\subsubsection{Comoment maps}
Theorem \ref{thmpb} defines a Hamiltonian lift of $\conf(M,g)$ to the supercotangent bundle $(\cM,d\a)$. We thus obtain a Hamiltonian action of $\conf(M,g)$ on $\cM$, as well as an equivariant momentum map $\cM\rightarrow\conf(M,g)^*$. By duality, this defines a comoment map which is an equivariant map or equivalently a Lie algebra morphism: $\conf(M,g)\rightarrow \cC^\infty(\cM)$.  
\begin{prop}\label{applimoment}
The even comoment map $\cJa:\conf(M,g)\rightarrow \cC^\infty(\cM)$, is a morphism of Lie algebras, whose expression is given by
\begin{equation}\label{applimomenta}
\medbox{\cJa_X=\langle \tilde{X},\alpha\rangle=p_iX^i+\frac{\hbar}{2\bi}\xi^j\xi^k\partial_{[k} X_{j]}.}
\end{equation} 
\end{prop}

\begin{rmk}\label{MomentSpin}
In the case of a flat manifold $(M,g)$, infinitesimal rotations are generated by the vector fields $X_{ij}=x_i\partial_j-x_j\partial_i$, and their  even comoments read 
$
\cJa_{X_{ij}}=(p_jx_i-p_ix_j)+2\frac{\hbar}{2\bi}\xi_i\xi_j
$.
The first term is the usual orbital momentum while the second one features the spin components $S_{ij}=\frac{\hbar}{\bi}\xi_i\xi_j$, as introduced in \cite{BMa77,Rav80}. 
\end{rmk}

\subsection{The supercotangent bundle of a conformally flat manifold}
We introduce a specific Darboux atlas on the supercotangent $\cM$, which is associated to every conformal atlas on $M$. This proves to be an efficient working tool in the last section. 
\subsubsection{Definition of a conformally flat manifold}

A conformal coordinate system $(x^i)$ on $(M,g)$, of signature $(p,q)$, is characterized by the fact that $g_{ij}=F\eta_{ij}$, where $\eta$ is a flat metric of signature $(p,q)$ and $F$ is a strictly positive smooth function. If there exists an atlas on $(M,g)$, where each chart is given by a conformal coordinate system, the manifold is said to be conformally flat. Then, locally, $\conf(M,g)\simeq\ro(p+1,q+1)$, and, in a conformal coordinate system, its generators are given by
\begin{eqnarray} \nonumber
X_i&=&\partial_i,\\ \nonumber
X_{ij}&=& x_i\partial_j- x_j\partial_i,\\ \label{GenConf}
X_0&=& x^i\partial_i,\\ \nonumber
\bar{X}_i&=&x_jx^j\partial_i - 2x_ix^j\partial_j,
\end{eqnarray}
for $i,j=1,\ldots,n=p+q$, the indices being lowered using the flat metric, $\eta$. 
The vector fields $X_{ij}$ generate the Lie algebra $\ro(p,q)$ of rotations. Together with the infinitesimal translations $X_i$ they form the Lie algebra of isometries  $\e(p,q)$, and $\ce(p,q)$ contains moreover the homothety $X_0$. The Hamiltonian lift of these vector fields can be obtained explicitly in the system of coordinates defined below from expressions (\ref{LdX0}) and (\ref{LdXi}), equating $\d$ to~$0$.

\subsubsection{Conformal Darboux coordinates on the supercotangent bundle}
We would like to lift any conformal atlas of $(M,g)$ to a conformal Darboux atlas on $(\cM,\om)$. This means to define Darboux coordinates $(x^i,\tp_i,\txi^i)$ from conformal coordinates $(x^i)$, such that the transition functions of $\cM$ are given by the Hamiltonian lift, defined in Theorem \ref{thmpb}, of those defining the conformal atlas of $M$. 
To construct such an atlas on $(\cM,\om)$, we resort to the even comoment map of the Hamiltonian lift of $\conf(M,g)$, defined in Proposition \ref{applimoment}, as well as to its odd comoment map, defined as follows. Let us recall that $d\b$ defines an odd symplectic structure on $\Pi TM$, and $\tilde{\b}=\b |\vol_g|^{-\frac{1}{n}}$ is preserved by the Hamiltonian lift $\tilde{X}$ of $X\in\conf(M,g)$, as shown by Equation (\ref{LXBeta}). Therefore, we can mimic the even case, and construct a $\conf(M,g)$-equivariant map $\cJb:\conf(M,g)\rightarrow \cC^\infty(\cM)\otimes\cF^{-\frac{1}{n}}$, whose expression is
\begin{equation}\label{applimomentb}
\cJb_X=\langle \tilde{X},\tilde{\beta}\rangle=\xi_i X^i |\vol_g|^{-\frac{1}{n}}.
\end{equation} 

\begin{prop}\label{PropConfDarboux}
Let $(M,g)$ be a conformally flat manifold and $(\cM,\om)$ be its supercotangent bundle. To every conformal coordinate system $(x^i)$, such that $g_{ij}=F\eta_{ij}$, there corresponds a~conformal Darboux coordinate system $(x^i,\tp_i,\txi^i)$ on $(\cM,\om)$, given by
\begin{equation}\label{tptxi}
\tp_i=\cJa_{\partial_i}= p_i-\frac{\hbar}{2\bi}\Ga_{ij}^k\xi^j\xi_k, \quad \text{and}\quad \txi_i=\cJb_{\partial_i}|\vol_x|^{\frac{1}{n}}= F^{-\half}\xi_i,
\end{equation}
with $\vol_x=dx^1\wedge\ldots\wedge dx^n$. Moreover, a conformal atlas of $(M,g)$ induces a conformal Darboux atlas on $(\cM,\om)$ via the latter correspondence. 
\end{prop}
\begin{proof}
As $(x^i)$ is a conformal coordinate system, the translation generators $(\partial_i)$ belong to $\conf(M,g)$ and the explicit formulas (\ref{applimomenta}) and (\ref{applimomentb}) of the comoments lead to (\ref{tptxi}). These are the Darboux coordinates of $\cM$ introduced in (\ref{CoordDarboux}), with the conformal change of frames $e^i_a=F^{-\half}\d^i_a$ and $\theta^a_i=F^{\half}\d^a_i$.
Furthermore, in the view of the $\conf(M,g)$-equivariance of the even and odd comoment maps, the transition functions are given by the Hamiltonian lift of the conformal transition functions of $(M,g)$.  
\end{proof}

\section{Geometric quantization of the supercotangent bundle and spinor geometry}\label{SectionQGcM}
 We perform, in this section, the geometric quantization of $(\cM,d\a)$. This relies on two ingredients: an extension of previous works \cite{Vor90,Tuy92} on  geometric quantization of supermanifolds over one point, and the choice of a polarization on $(\cM,d\a)$. Finally, we discuss the correspondence, set up by geometric quantization, between the conformal geometries of the supercotangent and the spinor bundles over~$M$. Notice that Bordemann has studied the deformation quantization of $(\cM,d\a)$ in \cite{Bor00}. 

\subsection{Geometric quantization scheme}
Geometric quantization has been developed by Kostant and Souriau as a geometrization of the orbit method of  Kirillov, and has been extended to the framework of supermanifolds \cite{Kos77,Tuy92,ENi96}.
We recall the Souriau's procedure of geometric quantization \cite{Sou70} adapted to the case of a symplectic supermanifold $(\cN,\sigma)$  in the special case of an exact symplectic $2$-form, $\sigma=d\varpi$. See \cite{Kos70} for the alternative approach of Kostant in terms of complex line bundles. 

Let $(\cN,d\varpi)$ be a symplectic supermanifold endowed with a polarization, see Definition~\ref{DefPola} below, and $\{\cdot,\cdot\}$ be its Poisson bracket. The purpose of geometric quantization is to construct a (quantum) representation space $\calH_{\GQ}$, a Lie subalgebra $\Obs$ of $\left(\Cinfty(\cN),\frac {\hbar}{\bi}\{\cdot,\cdot\}\right)$ and a morphism of Lie algebras $\cQ_{\GQ}:\Obs\rightarrow\left(\End(\calH_{\GQ}),[\cdot,\cdot]\right)$, where $[\cdot,\cdot]$ stands for the commutator. One of the aims of geometric quantization is to provide, as well, an Hilbert space structure on $\calH_{\GQ}$, such that the morphism $\cQ_{\GQ}$ takes its values in symmetric operators. We will not specifically discuss that point.

Let us first introduce prequantization. As the symplectic form is exact, the prequantum circle bundle is trivial, $\tilde{\cN}=\cN\times\bbS^1$, as well as the prequantum $1$-form $\tilde{\varpi}=\varpi+\hbar d\theta $, with $\theta$ the angular coordinate of $\bbS^1$. For every $f\in\Cinfty(\cN)$, we denote by $X_f^*$ the lift of the Hamiltonian vector field $X_f$ to $\tilde{\cN}$, satisfying $\big\langle X_f^*,\tilde{\varpi} \big\rangle=f$. It is called the quantum Hamiltonian vector field of $f$, namely
\begin{equation}\label{X_f^*}
 X_f^*=X_f+\frac{1}{\hbar}(f-\left\langle X_f,\varpi\right\rangle)\partial_{\theta}.
\end{equation}
As $X_f^*$ is determined by both conditions $L_{X_f^*}\tilde{\varpi}=0$ and $\big\langle X_f^*,\tilde{\varpi} \big\rangle=f$, this lift is a Lie algebra morphism. 
\begin{defi}
The prequantization of $(\cN,d\varpi)$ is the Lie algebra morphism
\begin{eqnarray}\nonumber
\cQ_{\PreQ}:\Cinfty(\cN)&\rightarrow & \End(\calH)\\ \label{cQPreQ}
 f &\mapsto & \frac{\hbar}{\bi}X_f^*,
\end{eqnarray}
where $\calH$ is the space of $\bbS^1$-equivariant smooth complex functions on $\tilde{\cN}$. 
\end{defi}
As a vector space, $\calH$ is isomorphic to the space $\Cinfty(\cN)$. The second step of geometric quantization consists in reducing the space of representation $\calH$ with the help of a polarization on $(\cN,d\varpi)$.
\begin{defi}\label{DefPola}
An admissible complex polarization on $(\cN,d\varpi)$ is a complex integrable Lagrangian distribution $\calP$ of $(\cN,d\varpi)$ such that the real distributions $\cE$ and $\D$, defined by $\cE=\calP\oplus\bar{\calP}\cap T\cN$ and $\D=\calP\cap\bar{\calP}\cap T\cN$, are respectively integrable and fibering. 
\end{defi}
Each vector field $X\in\calP$ admits a unique lift $\tilde{X}\in\Vect(\tilde{\cN})$ such that $\big\langle \tilde{X},\tilde{\varpi} \big\rangle=0$. One can choose, for the space of representation $\calH_{\GQ}$ of geometric quantization, the space of polarized functions
\begin{equation}
 \calH_{\pol}=\{\psi\in\calH\,|\; \tilde{X}\psi=0,\;\forall X\in\calP\}.
\end{equation}
The quantum action of $f\in\Cinfty(\cN)$ on $\calH_{\pol}$ is given by the prequantization (\ref{cQPreQ}). To obtain actual endomorphisms of $\calH_{\pol}$, we restrict ourselves to the space of quantizable functions, namely
\begin{equation}
 \Obs=\{f\in\Cinfty(\cN)\,|\;X_f^*  (\calH_{\pol})\subset  \calH_{\pol}\}.
\end{equation}
Geometric quantization can be further modified by considering a representation space $\calH_{\GQ}$ of half-forms rather than functions. In both cases the action of $X^*_f$ on $\calH_{\GQ}$ is given by the Lie derivative $L_{X^*_f}$.
 \begin{defi}
  The geometric quantization of $(\cN,d\varpi)$, endowed with the admissible polarization $\calP$, is the Lie algebra morphism
\begin{eqnarray}\nonumber
\cQ_{\GQ}:\Obs&\rightarrow & \End(\calH_{\GQ})\\ \label{GQgeneral}
 f &\mapsto & \frac{\hbar}{\bi}L_{X_f^*}.
\end{eqnarray}
 \end{defi}

\subsection{Geometric quantization and Clifford algebra representations}
We first apply geometric quantization to a one-point symplectic supermanifold $(\Pi V,\om)$. The $2$-form $\om$ is the differential of $\a=\frac{\hbar}{2\bi} g_{ij}\xi^id\xi^j$, where $g$ is a metric on $V$ and $(\xi^i)$ a dual basis. 

\subsubsection{Prequantization and representation on $\Lambda V^*$}

The prequantization of $(\Pi V,\om)$ has been investigated by B. Kostant \cite{Kos77} and leads to a representation of $\Cl(V^*,g^{-1})$ on $\Lambda V^*\simeq \calH$, the prequantization space. 
\begin{prop}
Let $(V,g)$ be a metric vector space and $(\Pi V,d\alpha)$ the associated symplectic supermanifold. The prequantization of $V^*\subset \Cinfty(\Pi V) $ induces a unique algebra morphism~$c$
\begin{equation}\label{PreQV}
\xymatrix @R=0.05in{
&& \bbCl(V^*,g^{-1}) \ar[dddr]^{c}&\\
&& &\\
&& &\\
V^* \ar@{^{(}->}[uuurr] \,\ar[rr]& & \End(\calH)\ar[r]^-{\simeq} &\End(\Lambda V^*) \\
v \ar@{|->}[rr] & & \cQ_{\PreQ}(\sqrt{2}v) \ar@{|->}[r] & \frac{1}{\sqrt{2}}(\ve(v)-2\iota(v)),}
\end{equation}
where $\ve(v)$ is the exterior product with $v$ and $\iota(v)$ the inner product with $g^{-1}(v)$.
\end{prop}
\begin{proof}
The Hamiltonian vector field of $f\in\Cinfty(\Pi V)$ is given by $X_f=(-1)^{|f|}\frac{\bi}{\hbar}g^{ij}\partial_{\xi^j}f\partial_{\xi^i}$, since $df=d\xi^i(\partial_{\xi^i}f)$. 
The general formula (\ref{X_f^*}) leads then to the quantum Hamiltonian vector field of $f$. In particular, the prequantization of $v=v_i\xi^i\in\Lambda ^1V^*$ is given by 
$
\cQ_{\PreQ}(v)=-\half v_i\left(\bi \xi^i\partial_{\theta}+2g^{ij}\partial_{\xi^j}\right).
$
Since $\Psi\in\calH$ has the form $\Psi(\xi,\theta)=e^{\bi \theta}\phi(\xi)$ with $\phi\in\Lambda V^*$, we obtain
$$
\cQ_{\PreQ}(\sqrt{2}v)\Psi(\xi,\theta)=e^{\bi \theta}c(v)\phi(\xi),
$$
where $c(v)=\frac{1}{\sqrt{2}}(\ve(v)-2\iota(v))$. 

Let $v,w\in V^*$. As prequantization is a Lie algebra morphism and $\frac{\hbar}{\bi}\{v,w\}=- g(v,w)$, the Clifford relations are satisfied: $c(v)c(w)+c(w)c(v)=-2g(v,w)$. Thanks to the universal property of the Clifford algebras, the map $c$ can be uniquely extended into an algebra morphism $\bbCl(V^*,g^{-1})\rightarrow \End(\Lambda V^*)$.
\end{proof}
\begin{rmk}
The prequantization of the symplectic supermanifold $(\Pi V,2d\alpha)$, composed with the map $v\mapsto 2v$ on $V^*$, leads to the canonical representation $c(v)=\ve(v)-\iota(v)$ of $\Cl(V^*,g^{-1})$ on $\Lambda V^*$ \cite{Kos77,Mic09}. 
\end{rmk}
\subsubsection{Geometric quantization and spinor representation}

The geometric quantization of $(\Pi V,\om)$ has been studied to a great extent by Voronov \cite{Vor90}, in the case of an Euclidean metric $g$, while Tuynman \cite{Tuy92} has treated the case of a metric~$g$ of signature $(p,p)$. As in the above-mentioned articles, we will suppose that $V$ is of even dimension $2n$, the novelty residing in the arbitrary signature $(p,q)$ of the metric $g$. As a consequence, we will have to deal with a mixed real-complex polarization, instead of a K\"ahler~\cite{Vor90} or real \cite{Tuy92} one.  We assume that $p\geq q$ and we denote by $(\eta_{ij})=\mathbb{I}_p\otimes-\mathbb{I}_q$ the matrix of~$g$ in an orthonormal basis. 

To perform  geometric quantization of $(\Pi V,d\a)$ we need a polarization. In this context, this means a maximal isotropic subspace $P$ of $V\otimes\bbC$ for the complex linear extension of $g$. As in the general setting,  we define the spaces $E=(P\oplus\bar{P})\cap V$ and $D=P\cap\bar{P}\cap V$. In the case of 	an Euclidean metric they are trivial, i.e$.$ $E=V$ and $D=\{0\}$. Then, $\bar{P}$ is a polarization and a supplementary space of $P$, that plays a crucial role in the geometric quantization of~$\Pi V$~\cite{Vor90}. In the general case, we require an analog of $\bar{P}$, which relies on a choice.
\begin{defi}
Two polarizations $P$ and $\btP$ are said to be conjugate if $P\oplus\btP=V\otimes\bbC$ and $\btP$ is of the form $\btP=(\btP\cap\bar{P})\,\oplus\,(\hat{D}\otimes\bbC)$, with $\hat{D}$ a real vector subspace of $V$. 
\end{defi}
Starting from a polarization $P$, the construction of a conjugate polarization $\btP$ amounts to choosing a real supplementary space $\hat{D}$ of $E$, defined above. The space $\btP$ is then given by $(\bar{P}\cap\ker\om(\hat{D},\cdot))\oplus(\hat{D}\otimes\bbC)$, where $\ker\om(\hat{D},\cdot)$ is the intersection of the kernels of $\om(u,\cdot)$,  for all $u\in\hat{D}$.
\begin{lem}\label{LemHpol}
Let $P$ and $\btP$ be two conjugate polarizations. The space $\calH_{\pol}$ of polarized functions on $\Pi V$ with respect to the polarization $\btP$, identifies as a vector space to $\Cinfty(\Pi P)=\Lambda P^*$.
\end{lem}
\begin{proof}
We determine explicitly the space of polarized functions from the prequantum $1$-form.
To that end, we introduce K\"alher-like coordinates $(\zeta^a,\bzeta^a)_{a=1,\ldots,n}$ such that: $\om=\frac{\hbar}{\bi}\d_{ab}d\zeta^a\wedge d\bzeta^b$, their Hamiltonian vector fields generate $P=\big\langle\partial_{\zeta^a} \big\rangle$ and $\btP=\big\langle\partial_{\bzeta^a} \big\rangle$, as well as $D$ and $\hat{D}$ for {$a\geq \frac{p-q}{2}+1$}, and are conjugate $\bzeta^a=\bar{\zeta}^a$ if $a\leq \frac{p-q}{2}$.

We can then define Darboux coordinates $(\xi^i)_{i=1,\ldots,2n}$ on $\Pi V$, by $\xi^i=\frac{1}{\sqrt{2}}(\zeta^i+\bzeta^i)$ if $i\leq n$, $\xi^i=\frac{\bi}{\sqrt{2}}(\zeta^{i-n}-\bzeta^{i-n})$ if $n<i\leq p$ and $\xi^i=\frac{1}{\sqrt{2}}(\zeta^{i-n}-\bzeta^{i-n})$ if $p<i\leq 2n$. The $1$-form $\a$ has the standard expression $\a=\frac{\hbar}{2\bi}\eta_{ij}\xi^id\xi^j$ in these coordinates, and by substitution we end with the following expression for the prequantum $1$-form, 
\begin{equation}\label{tildealphaV}
\tilde{\a }=\frac{\hbar}{2\bi} \d_{ab}\left(2\bzeta^ad\zeta^b+ d(\zeta^a\bzeta^b)\right)+\hbar d\theta. 
\end{equation}
It allows us to compute the lift $\tilde{X}_{\zeta^a}=X_{\zeta^a}-\frac{1}{2\hbar}\zeta^a\partial_{\theta}$ of the generators of the polarization $\btP$ to the prequantum bundle, and to determine explicitly the space of polarized functions 
\begin{equation}\label{PsiHpol}
\calH_{\pol}=\{\Psi\in\cC^\infty_\bbC(\Pi V\times\bbS^1)\,|\;\Psi(\zeta,\bzeta,\theta)=e^{\bi\theta}e^{\half\delta_{ab}\zeta^a\bzeta^b}\psi(\zeta), \;\text{with}\;\psi\in\Cinfty(\Pi P)\}.
\end{equation}
Hence, we have $\calH_{\pol}\simeq\Cinfty(\Pi P)$. 
\end{proof}

We choose $\calH_{\GQ}=\calH_{\pol}\otimes (\Ber(P))^\half$ as representation space, where $(\Ber(P))^\half$ is the one dimensional vector space of half-forms on $P$ \cite{Vor90}, the action of $M\in\GL_+(P)$ being given by the square root of its Berezinian $(\Ber(M))^{\half}$, equal to $(\det(M))^{-\half}$ \cite{Lei80}. That choice of $\calH_{\GQ}$ will prove necessary, in order that the geometric quantization mapping coincides with Weyl quantization. This is reminiscent of the case of the cotangent bundle $T^*M$, where the geometric quantization mapping coincides with Weyl quantization for a quantum space of half-densities on $M$ \cite{Bla72,Kos74}. 

\begin{thm}\label{ThmGQPV}
Let $P$ and $\btP$ be two conjugate polarizations and $\g$ be defined by (\ref{gLvCLv}). The geometric quantization of $(\Pi V,d\a)$, endowed with the polarization $\btP$, is defined on $\Obs=(\Lambda^0V^*\oplus\Lambda^1V^*)\cdot \Lambda P^*$, and induces the  algebra isomorphism $\varrho$,
\begin{equation}\label{QGQW}
\xymatrix{
\Lambda V^*\otimes\bbC \ar[rr]^\g&& \bbCl(V^*,g^{-1}) \ar[dr]^{\varrho}&\\
\Obs \ar@{^{(}->}[u] \,\ar[rr]^{\cQ_{\GQ}} && \End(\calH_{\GQ})\ar[r]^-{\simeq} &\End(\Lambda P^*\otimes\Ber(P)^\half),
}
\end{equation}
which turns the vector space $\calH_{\GQ}\simeq\Lambda P^*\otimes\Ber(P)^\half$ into the module of spinor of $\bbCl(V^*,g^{-1})$.
Moreover, each map of the diagram (\ref{QGQW}) is equivariant under the action of $\GL(P^*)$.
\end{thm}
\begin{proof}
We keep the notation introduced in the proof of Lemma \ref{LemHpol}. 

Firstly, we determine $\Obs$ and give an explicit expression of $\cQ_{\GQ}$. Resorting to the coordinate expression (\ref{tildealphaV}) of $\tilde{\a}$, a direct computation gives the quantum Hamiltonian vector field $X_f^*$ of $f\in\cC^\infty_\bbC(\Pi V)$. It acts on $\Psi\in\calH_{\pol}$, obtained in (\ref{PsiHpol}), according to 
\begin{equation}\label{XfLV}
\frac{\hbar}{\bi}X_f^*\Psi(\zeta,\bzeta,\theta)=e^{i\theta}e^{\half\delta_{ab}\zeta^a\bzeta^b}\left( 
(-1)^{|f|}\d^{ab}\partial_{\bzeta^a}f\partial_{\zeta^b}
+\big[f-\bzeta^a\partial_{\bzeta^a}f\big]\right)\psi(\zeta),
\end{equation}
with $\psi\in\Lambda P^*$. Clearly, if $f$ is of degree $2$ or more in the $\bzeta^a$ coordinates, then the operator $\frac{\hbar}{\bi}X_f^*$ does not preserve the space of polarized functions. We conclude that $f\in\Obs$ is of the form $f(\zeta,\bzeta)=\bzeta^a A_a(\zeta)+B(\zeta)$. With the help of (\ref{XfLV}), we see that such a function acts on the half-form $\nu\in(\Ber P)^\half$ by
$
L_{X_f^*}\,\nu=\left(-\half\d^{ab}\partial_{\bzeta^a}\partial_{\zeta^b}f\right)\nu
$.
 The geometric quantization operator $\cQ_{\GQ}$ reads, hence, 
\begin{equation}\label{QGObsLP}
 \cQ_{\GQ}(f)(\Psi\otimes\nu)= e^{i\theta}e^{\half\delta_{ab}\zeta^a\bzeta^b}\left( 
(-1)^{|f|}A^a(\zeta)\partial_{\zeta^a}-\half\partial_{\zeta^a}A^a(\zeta)
+B(\zeta)\right)\psi(\zeta)\otimes\nu.
\end{equation}

Secondly, we define the map $\varrho$ assuming that diagram (\ref{QGQW}) restricted to $V^*$ commutes. For all $u,v\in V^*$, we have $\frac{\hbar}{\bi}\{v,w\}=-g(v,w)$, and since geometric quantization is a Lie algebra morphism, the elements $\varrho(\g(v)),\varrho(\g(w))$ satisfy the Clifford relations. The map $\varrho$ can therefore be uniquely extended to an algebra morphism ${\varrho:\bbCl(V^*,g^{-1})\rightarrow\End(\Lambda P^*\otimes\Ber(P)^\half)}$, with $\varrho(\g(\zeta^a))=\zeta^a$ and $\varrho(\g(\bzeta^a))=-\partial_{\zeta^a}$. Since this morphism has trivial kernel and the two algebras have the same dimension over $\bbC$, $\varrho$ is an isomorphism. 

Thirdly, we have to prove that this extension of $\varrho$ coincides with geometric quantization on all $\Obs$. Let $\bzeta^a A_a(\zeta)+B(\zeta)\in\Obs$. Denoting by $\star$ the Moyal product, defined in Section~\ref{SecPiV}, we have $\bzeta^a\star A_a(\zeta)=\bzeta^aA_a(\zeta)-\half\partial_{\zeta^a}A^a(\zeta)$, and by Proposition \ref{gLie}, we are left with
$$
\varrho(\g(\bzeta^a A_a(\zeta)+B(\zeta)))=(-1)^{|f|}A^a(\zeta)\partial_{\zeta^a}-\half\partial_{\zeta^a}A^a(\zeta)+B(\zeta),
$$
which coincides with (\ref{QGObsLP}), proving the commutativity of the diagram (\ref{QGQW}).

Fourthly, we study the equivariance under the action of $\GL(P^*)$.
We endow the subspaces of $\Lambda V^*\otimes\bbC$ with the Poisson bracket coming from $d\a$. Then, we get that each map of the diagram (\ref{QGQW}) is equivariant under the action of $\GL(P^*)$, since: $\cQ_{\GQ}$ is a Lie algebra morphism and $\Obs$ contains $P^*\otimes \hat{P}^*\simeq\gl(P^*)$, the isomorphism $\calH_{\GQ}\simeq\Lambda P^*\otimes\Ber(P)^\half$ consists in suppressing the phase term $e^{i\theta}e^{\half\delta_{ab}\zeta^a\bzeta^b}$ (see (\ref{XfLV})) which is invariant under $P^*\otimes \hat{P}^*$, and Proposition \ref{gLie} shows that the map $\g$ is equivariant under $\Lambda^2V^*\otimes\bbC\supset P^*\otimes \hat{P}^*$. 
\end{proof}
The composition map $\varrho\circ\g$ is called the Weyl quantization, by analogy with the even case, and coincides with the usually defined spinor representation. Theorem \ref{ThmGQPV} together with Proposition \ref{gLie} lead to the following corollary. 
\begin{cor}\label{CorQGQW}
Geometric quantization of $(\Pi V,\om)$ coincides on its definition space with Weyl quantization. Moreover, it can be extended to $\Lambda^0 V\oplus\Lambda^1 V\oplus\Lambda^2 V$ as a Lie algebra morphism. 
\end{cor}
 
\subsection{From $(\cM,\om)$ to the spinor bundle of $(M,g)$}

Locally, via Darboux coordinates, the supercotangent bundle is symplectomorphic to the product of the two symplectic manifolds $T^*\bbR^n$ and $\Pi\bbR^n$. Consequently, the geometric quantization of $\cM$ arises, locally, directly as the product of those of $T^*\bbR^n$ and $\Pi\bbR^n$, and, so, we do not need to perform the prequantization of $\cM$ \cite{Rot90}. An admissible polarization on $\cM$ allows then to define globally a quantum representation space, which will, as we shall prove, be the space of sections of the spinor bundle over $M$. 

\subsubsection{Geometric quantization of the flat supercotangent bundle}

Geometric quantization of the supercotangent bundle of $(\bbR^n,\eta)$, with $\eta$ the flat metric of signature $(p,q)$, is simply given by merging those of $T^*\bbR^n$ and $\Pi\bbR^n$. 
We have already studied the geometric quantization of $\Pi\bbR^n$, and that of $T^*\bbR^n$ is canonically defined thanks to its vertical polarization $\left\langle \partial_{p_i}\right\rangle$. We choose as quantum representation space the space of polarized functions, isomorphic to $\Cinfty(\bbR^n)$, rather than polarized half-densities, so that $\cQ_{\GQ}$ coincides with normal ordering, rather than Weyl quantization. 
Just as on $\Pi \bbR^n$, there is no canonical polarization on the supercotangent bundle of $(\bbR^n,\eta)$. Nevertheless, to every polarization on~$\Pi\bbR^n$ there is an associated one on the supercotangent bundle, given by its direct sum with the vertical one of $T^*\bbR^n$. 
\begin{prop}\label{PropQGSRn}
Let $P,\btP$ be two conjugate polarizations on $(\Pi\bbR^n,\eta)$. The geometric quantization of the supercotangent bundle $T^*\bbR^n\times\Pi\bbR^n$, endowed with  the polarization $\left\langle \partial_{p_i}\right\rangle\times\btP$, is defined by the morphism,
\begin{equation}\label{QGSRn}
\begin{array}{ccccc}
 \Obs &\rightarrow & \End(\calH_{\GQ})&\simeq& \End(\Cinfty(\bbR^n))\otimes\bbCl(\bbR^n,\eta))\\ \nonumber
f=A^ip_i+B_j\xi^j+C & \mapsto &\;\frac{\hbar}{\bi}L_{X_f^*}&\mapsto& \frac{\hbar}{\bi}A^i\partial_i+B_j\frac{\g^j}{\sqrt{2}}+C,
\end{array}
\end{equation}
where $A_i,B^j,C$ are functions on $\bbR^n\times\Pi P$ and $\calH_{\GQ}\simeq\Cinfty(\bbR^n\times\Pi P)\otimes\Ber(\Pi P)^\half$.
\end{prop} 

\subsubsection{Polarization on $\cM$ and the spinor bundle}

As for the supercotangent bundle of $(\bbR^n,\eta)$, we are looking for a polarization on $\cM$ which projects as a polarization on both $T^*M$ and $\Pi T_xM$, for every $x\in M$. We choose the vertical polarization on $T^*M$. We have to complete it with a maximal isotropic distribution of $\Pi T^\bbC M$ for the complex linear extension of $g$. Such a distribution is provided by the notion of $N$-structure over $M$, introduced by Nurowski and Trautmann.
\begin{defi} \cite{NTr02}\label{N-structure}
A $N$-structure on a pseudo-Riemannian manifold $(M,g)$ of even dimension is a complex subbundle $N$ of $T^\bbC M$, whose fibers are maximal isotropic for the $\bbC$-linear extension of the metric $g$. 
\end{defi}
\begin{rmk}
 In the case of a Riemannian metric, a $N$-structure on $(M,g)$ is equivalent to an almost complex structure on $M$, given by $J(v)=\bi v$ and $J(\bar{v})=-\bi \bar{v}$ for $v\in N$. 
\end{rmk}
Let us suppose from now on that $(M,g)$ is of even dimension and admits a $N$-structure, which is a non-trivial topological requirement on $M$. 
We naturally choose as polarization on $\cM$ the distribution $\cPN$ generated by the vertical vector fields of $T^*M\times_M \Pi N$ over $M$. This polarization is trivially admissible, fibered over $\Pi T^\bbC M/N$ and its projections define polarizations on $T^*M$ and $\Pi T^\bbC_xM$ for each $x\in M$.
We denote by $\calH_{\pol}$ the space of polarized functions of~$(\cM,\om)$, equipped with the polarization $\cPN$. 

We choose, as quantum representation space, $\calH_{\GQ}=\calH_{\pol}\otimes\Ga(\det(T^\bbC M/N)^{-\half})$, where $\det(E)$ denotes the higher exterior power $\Lambda^{top}E^*$. The square root of this line bundle exists if and only if the first Chern class $c_1(M)$ of the manifold $M$ is divisible by $2$ \cite{Vor90}.
\begin{thm}\label{thmSpinorB}
 Let $N$ be a $N$-structure on $(M,g)$, $c_1(M)=0 \,\mathrm{mod}\, 2H^2(M;\bbZ)$ and let $\g$ be defined as in Corollary \ref{GradCliff}. The geometric quantization of $(\cM,\om)$, endowed with the polarization $\cPN$, yields a Lie algebra morphism $\cQ_{\GQ}$, given by (\ref{QGSRn}) in Darboux coordinates, and defined on functions which are polynomial of degree at most $1$ in $p$ and $2$ in $\xi$. It induces the algebra isomorphism~$\varrho$,
\begin{equation}\label{MQGQW}
\xymatrix{
&& \Ga(\bbCl(M,g)) \ar[drr]^{\varrho}&&\\
\Ga(\Lambda^1T^*M) \ar@{^{(}->}[urr] \,\ar[rr]^{\cQ_{\GQ}} &&  \End(\calH_{\GQ})\ar[rr]^{\simeq} &&\End(\Ga(\sS)),
}
\end{equation}
which turns the vector bundle $\sS$, s.t. $\Ga(\sS)\simeq\calH_{\GQ}$, into the spinor bundle of $(M,g)$.
\end{thm}
\begin{proof}
We keep the notation of Theorem \ref{ThmGQPV} and Proposition \ref{PropQGSRn}. 

The Newlander-Nirenberg theorem ensures the existence of local coordinates $(x^i,\tp_i,\zeta^a,\bzeta^a)$ such that $\om=d\tp_i\wedge dx^i+\frac{\hbar}{\bi}\d_{ab}d\zeta^a\wedge d\bzeta^b$ and $\cPN=\big\langle \partial_{\tp_i},\partial_{\bzeta^a}\big\rangle$. Those coordinates provide a local symplectomorphism between the supermanifolds $\cM$ and $T^*\bbR^n\times\Pi\bbR^n$, which sends one polarization to the other. As $\Ber=\det^{-1}$ on odd vector spaces, they establish, as well, a local isomorphism between their quantum spaces of representation. 
Hence, the geometric quantization of $\cM$ is given locally by Proposition \ref{PropQGSRn} and we readily deduce that $\cQ_{\GQ}$ is given by (\ref{QGSRn}) in Darboux coordinates.

Since the diagram (\ref{QGQW}) is  $\GL(P^*)$-equivariant, the principal bundle of complex linear frames $\GL(T^\bbC M/N)$ allows us to geometrize it, and thus to obtain the diagram (\ref{MQGQW}). In particular, $\sS$ is the associated bundle $T^\bbC M/N\otimes\det(T^\bbC M/N)^{-\half}$ and satisfies $\End(\sS)\simeq\bbCl(M,g)$, i.e$.$ $\sS$ is the spinor bundle of $M$. 
 The content of Corollary \ref{CorQGQW} can also be geometrized, which provides the extension of $\cQ_{\GQ}$ to functions of degree $2$ in $\txi$, and then to the natural coordinates $p_i=\tp_i-\frac{\hbar}{2\bi}\om_{bi}^a\txi_a\txi^b$, see (\ref{CoordDarboux}). 
\end{proof}
This theorem exhibits a new construction of the spinor bundle $\sS$, which turns out to be explicitly given by $\Lambda \btN\otimes\det(\btN)^{-\half}$ if there exists a subbundle $\btN$ of $\Pi T^\bbC M$ such that $N_x$ and~$\btN_x$ are two conjugate polarizations on $\Pi T^\bbC_xM$. In the case of a Riemannian metric~$g$, such a subbundle is provided by $\bar{N}$, and we recover the well-known construction of the spinor bundle of an almost-complex spin manifold \cite{Hit74,LMi89}. Let us provide some details. As already mentioned, a $N$-structure on a Riemannian manifold $(M,g)$ is equivalent to an almost complex structure on $M$, and $N$ is then the  holomorphic tangent bundle $T^{1,0}M$. This implies the existence of a canonical $\spin^c$ structure on $M$. Next, the existence of the square root of $\det (\btN)$ corresponds to the vanishing of the first Chern class modulo $2$, which is precisely the condition to extract a spin structure from a $\spin^c$ one. The associated spinor bundle is $\Lambda T^{0,1}M\otimes K^\half$ \cite{Hit74,LMi89}, with $K$ the canonical bundle. Since $K\simeq \Lambda^{top}(T^{1,0}M)^*$, we finally have $K^\half\simeq\det (\btN)^{-\half}$ and the two constructions coincide indeed. 

In the generic case of a pseudo-Riemannian metric, we do not know if the  conditions required for $M$ in Theorem \ref{thmSpinorB} imply the existence of a spin structure on $M$.


\subsection{The Lie derivative of spinors}
We just have constructed the spinor bundle of $(M,g)$ out of the supercotangent bundle $(\cM,\om)$. We now go further and obtain the covariant derivative and Lie derivative on the spinor bundle of $M$ by means of the quantization of Hamiltonian actions on $T^*M$ and $\cM$. 
So, we get spinor geometry from the symplectic geometry of the supercotangent bundle via geometric quantization. This yields a new interpretation of the Lie derivative of spinors, which has attracted much attention \cite{BGa92,GMa05,PWi08} since its introduction by Kosmann \cite{Kos72}. From now on, we denote by $\sS$ the spinor bundle of $M$ and we identify $\Ga(\sS)$ with $\calH_{\GQ}$ when it exists.

Recall that $\cJa$, introduced in Proposition \ref{applimoment}, is the comoment map associated with the Hamiltonian action of $\conf(M,g)$ on the supercotangent bundle $\cM$. We denote by $J$ the trivial lift to $\cM$ of the comoment map associated to the Hamiltonian action of $\Vect(M)$ on~$T^*M$, whose expression is $X\mapsto J_X=p_iX^i$. 
\begin{thm}\label{thmSpinorL}
If the hypothesis of Theorem \ref{thmSpinorB} holds, the geometric quantization mapping $\cQ_{\GQ}$ of $\cM$ establishes the correspondences
\begin{equation}\label{EqNablaS}
\medbox{\forall X\in\Vect(M),\quad\cQ_{\GQ}(J_X)=\frac{\hbar}{\bi}\nabla_X,} 
\end{equation}
where $\nabla_X$ is the covariant derivative of spinors along $X$, and
\begin{equation}\label{EqsL}
\medbox{\forall X\in\conf(M,g),\quad\cQ_{\GQ}(\cJa_X)=\frac{\hbar}{\bi}\sL_X,} 
\end{equation}
where $\sL_X$ is the Lie derivative of spinors along $X$, proposed by Kosmann \cite{Kos72}. Both correspondences still hold if $M$ is a spin manifold.
\end{thm}
\begin{proof}
It suffices to work in local Darboux coordinates to prove this theorem. First, let $X\in\Vect(M)$, then $J_X=p_iX^i$ and, since $\tp_i=p_i+\frac{\hbar}{2\bi}\om_{bi}^a\txi_a\txi^b$ ( see (\ref{CoordDarboux})), we have 
\begin{equation}
 \cQ_{\GQ}(J_X)=\frac{\hbar}{\bi}\left(X^i\partial_i+\frac{1}{4}\om_{bi}^a\tg_a\tg^b \right),
\end{equation}
where $\tg^i=\g(\txi^i)$, the map $\g$ being defined in Corollary \ref{GradCliff}. This is precisely the expression of the covariant derivative of spinors along $X$. Secondly, we restrict to {$X\in\conf(M,g)$}, admitting as comoment $\cJa_X$, given in (\ref{applimomenta}). Its quantization leads to
\begin{equation}\label{sLX}
 \cQ_{\GQ}(\cJa_X)=\frac{\hbar}{\bi}\left(X^i\nabla_i+\frac{1}{4}(\partial_{[k} X_{j]} )\g^j\g^k \right),
\end{equation}
which is precisely the expression of the Lie derivative of spinors, as introduced by Kosmann.

On the one hand, the geometric quantization of the supercotangent bundle $(\cM,\om)$ is ever well-defined locally. On the other hand, the spinor bundle, the covariant derivative and the Lie derivative of spinors are global if there is a spin structure on $M$. Hence, both correspondences (\ref{EqNablaS}) and (\ref{EqsL}) generalize to the case where $M$ is endowed with a spin structure.
\end{proof}
A still active research activity is devoted to a better understanding of the Lie derivative of spinors \cite{BGa92,GMa05}, which is not a canonical geometric object, for a given spin structure. It is defined as a Lie algebra morphism $\sL:\fkg\rightarrow \End(\Ga(\sS))$, with source a Lie subalgebra of~$\Vect(M)$ and target the derivations of the spinor fields. Thus, a Lie derivative of spinors can be written as $\sL_X=\nabla_X+A(X)$, where $A$ is a section of the Clifford bundle depending linearly on $X\in\fkg$. We can further specify $A$ to be a section of the subbundle of Lie algebras $\spin(M)$, and then a Lie derivative of spinors is necessarily of the form
\begin{equation}\label{GeneralLieDerivative}
 \sL_X=\nabla_X+Y_{ij}\g^i\g^j,
\end{equation}
 with $Y$ a skew-symmetric tensor depending linearly on $X$. This expression coincides with that of Godina and Matteucci \cite{GMa05}, in their approach of general Lie derivatives on gauge-natural bundles.
\begin{prop}\label{ThmLieS}
 Let $\fkg\subset\Vect(M)$ be a Lie algebra, and restrict the Lie derivatives of spinors to those of the form (\ref{GeneralLieDerivative}). We then have the following correspondence
\begin{eqnarray*}
\Big( \text{ Hamiltonian lift } \rho:\fkg\rightarrow\Vect(\cM)\;\Big)& \Longleftrightarrow &  \Big(\text{ Lie derivative of spinors } \sL:\fkg\rightarrow \End(\Ga(\sS))\;\Big),
\end{eqnarray*}
as long as there is a spin structure on $M$. This is given explicitly by $\frac{\hbar}{\bi}\sL_X=\cQ_{\GQ}(\left\langle \rho(X),\a \right\rangle )$ and $\rho(X)=-\om^{-1}(d\cQ_{\GQ}^{-1}(\frac{\hbar}{\bi}\sL_X))$ if the assumptions of Theorem \ref{thmSpinorB} hold.
\end{prop}
\begin{proof}
We suppose that the hypothesis of Theorem \ref{thmSpinorB} hold, insuring the existence of $\cQ_{\GQ}$.
 
 Let $\rho:\fkg\rightarrow\Vect(\cM)$ be a Hamiltonian lift, and as such this is determined by the Lemma \ref{LXaLem}. Consequently, the comoment map, defined by $X\mapsto \left\langle \rho(X),\a \right\rangle$, takes its values in the space of quantizable functions of $\cM$. As the comoment map and $\cQ_{\GQ}$ are both Lie algebra morphisms, the same holds for $\frac{\hbar}{\bi}\sL:X\mapsto \cQ_{\GQ}(\left\langle \rho(X),\a \right\rangle )$. Thanks to Lemma \ref{LXaLem}, we can compute $\sL_X$, which retains the expression (\ref{GeneralLieDerivative}).

We suppose now the existence of a Lie derivative of spinor fields on $\fkg$ of the form (\ref{GeneralLieDerivative}). As a consequence, $\sL_X$ is in the image of $\cQ_{\GQ}$, and taking its inverse we get a function on $\cM$ whose Hamiltonian vector field is a lift of $X$ to $\cM$. This is the sought Lie algebra morphism~$\rho$. 

For the general case, geometric quantization gives a correspondence between local objects and their global meaning arise from the spin structure on $M$.
\end{proof}
\begin{rmk}
In the view of the correspondence stated by Proposition \ref{ThmLieS}, the choice made by Kosmann can be translated in symplectic terms as the choice of the Hamiltonian lift which preserves the direction of $\b$. This gives a new interpretation of the Lie derivative of spinors introduced by Kosmann.
\end{rmk} 
\begin{rmk}
The general expression (\ref{GeneralLieDerivative}) of a potential Lie derivative of spinors has no reason to give rise to Lie algebras morphism. So, the question of the uniqueness of the Lie derivative of spinors is still an open problem, and we hope this new framework might help to address it.
\end{rmk}

\section{Spinor differential operators and their symbols}

We study in this section the space $\Dslm$ of differential operators acting on spinor densities and the space of their Hamiltonian symbols $\cS^\d[\xi]$, endowed with a $\conf(M,g)$-module structure, via the classical and quantum $\conf(M,g)$-actions previously defined. In the conformally flat case, we compare those modules to the one of tensorial symbols $\cT^\d[\xi]$, and deduce a classification of their conformal invariants.
 In all this section, we suppose that $M$ is endowed with a spin structure, we denote by $\sS$ its spinor bundle and by $n$ its dimension.  

\subsection{Definition of three $\conf(M,g)$-modules}

Recall that $\D(M,\sS)$ is the algebra of spinor differential operators, and $\cS(M)[\xi]$ the associated bigraded algebra of symbols for the natural bifiltration (see Section \ref{Sec212}). The Lie derivative of spinors defines by adjunction a $\conf(M,g)$-action on $\D(M,\sS)$, and the Hamiltonian lift from $M$ to its supercotangent bundle $\cM$ also defines a $\conf(M,g)$-action on $\cS(M)[\xi]$. 
We will introduce tensor densities, which deform those actions and lead to the modules of differential operators $\Dslm$ and of Hamiltonian symbols $\cS^\d[\xi]$. From their natural bifiltration, we deduce on both modules a natural and a Hamiltonian filtration. For the latter, we show that the space of Hamiltonian symbols $\cS^\d[\xi]$ is indeed isomorphic to the graded module associated to $\Dslm$, and, for $\l=\m$, the Poisson bracket on $\cM$ turns $\cS^0[\xi]$ into the graded Poisson algebra associated to $\Dslm$. We also introduce a third module $\cT^\d[\xi]$ of tensorial nature which stems from the natural filtrations above.

\subsubsection*{Preliminary: the $\Vect(M)$-module of tensor densities $\cF^\l$}

A tensor density of weight $\l$ is a section of the line bundle $|\Lambda^nT^*M|^{\otimes\l}$. We denote by $\cF^\l$  the space of tensor densities if there is no ambiguity on the chosen manifold $M$. This space is naturally endowed with a $\Vect(M)$-module structure. 
In a coordinate system $(x^i)$ of $M$, there exists a local $1$-density $|\vol_x|=|dx^1\wedge\cdots\wedge dx^n|$. The $\l$-densities write then locally as $f|\vol_x|^\l$ with $f\in\Cinfty(M)$, and the $\Vect(M)$-module $\cF^\l$ is locally identified with the space of functions $\Cinfty(M)$, endowed with the action $\ell^\l$ of~$\Vect(M)$, namely
\begin{equation}\label{Ll}
\Ll=X^i\partial_i+\l(\partial_i X^i),
\end{equation}
for all $X\in\Vect(M)$.
If $M$ is endowed with a metric $g$, there is a canonical $1$-density, given by
$
|\vol_g|=\sqrt{|\det (g_{ij})|}\,|dx^1\wedge\cdots\wedge dx^n|.
$ 

\subsubsection{The module of tensorial symbols $\cT^\d[\xi]$}

Let us recall that the algebra of spinor symbols $\cS(M)[\xi]$ is isomorphic to the tensor algebra $\Ga(\cS TM\otimes \Lambda T^*M)$, see (\ref{Tenseur_Poly}), and admits therefore a natural action of $\Vect(M)$. This allows us to introduce a $\Vect(M)$-module structure on $\cS(M)[\xi]$, preserving its bigradation given by the degrees in $p$ and $\xi$. We denote by $\Sigma=\xi^i\partial_{\xi^i}$ the odd Euler operator.
\begin{defi}
Let $(x^i,p_i,\xi^i)$ be natural coordinates on $\cM$. The module of tensorial symbols is the space
{$
\cT^\d[\xi]=\bigoplus_{k=0}^\infty\bigoplus_{\k=0}^n\cS_{k,\k}(M)[\xi]\otimes\cF^{\d-\frac{\k}{n}},
$}
endowed with the natural action of~$\Vect(M)$,
\begin{equation}\label{Lt}
\Lt=X^i\partial_i-p_j\partial_iX^j\partial_{p_i}+\xi^i\partial_iX^j\partial_{\xi^j}+\left(\d-\frac{\Sigma}{n}\right)\partial_iX^i.
\end{equation}
\end{defi}
\begin{rmk}
The action (\ref{Lt}) of $X\in\Vect(M)$ on $T^\d[\xi]$ depends explicitly on the degree in the odd fiber variables. This choice of action will prove necessary later on.
\end{rmk}

\subsubsection{The module of Hamiltonian symbols $\cS^\d[\xi]$}

Theorem \ref{thmpb} proves that the algebra of spinor symbols $\cS(M)[\xi]$ carries a $\conf(M,g)$-action, provided by the Hamiltonian lift $X\mapsto\tilde{X}$ of $\conf(M,g)$ to $\cM$. This action defines the structure of $\conf(M,g)$-module on $\cS(M)[\xi]\subset\cC^\infty_\bbC(\cM)$, as a classical space of observables.

\begin{defi}
The module of Hamiltonian symbols is the space 
$
\cS^\d[\xi]=\cS(M)[\xi]\otimes\cF^\d
$ 
endowed with the Hamiltonian action of $\conf(M,g)$,
\begin{equation}\label{Ld}
\Ld=\tilde{X}\otimes\Id+\Id\otimes\ell_X^\d.
\end{equation}
\end{defi}
In accordance with (\ref{Tenseur_Poly}), we denote by $\cS_{k,\k}^\d[\xi]=\cS_{k,\k}(M)[\xi]\otimes\cF^\d$ the subspace of Hamiltonian symbols of degrees $k$ in $p$ and $\k$ in $\xi$, where  $(x^i,p_i,\xi^i)$ are natural coordinates. From the expression (\ref{tildeX}) of $\tilde{X}$ in a Darboux coordinates system, these subspaces are clearly not preserved by the action (\ref{Ld}). We rather have the following results.
\begin{defi}\label{DefNHFiltre}
The $\conf(M,g)$-module $\cS^\d[\xi]$ admits a natural filtration by the degree in~$p$: $\cS_0^\d[\xi]\subset\cS_1^\d[\xi]\subset\cdots\subset\cS_k^\d[\xi]\subset\cdots$, where $\cS_k^\d[\xi]=\bigoplus_{j=0}^k\bigoplus_{\k=0}^n\cS_{j,\k}^\d[\xi]$. Moreover, $\cS^\d[\xi]$ turns into a graded $\conf(M,g)$-module $\cS^\d[\xi]=\bigoplus_{k=0}^\infty\cS^\d_{(\frac{k}{2})}[\xi]$, for the Hamiltonian gradation defined by $\cS^\d_{(\frac{k}{2})}[\xi]=\bigoplus_{2j+\k=k}\cS_{j,\k}^\d[\xi]$.   
\end{defi} 
Geometrically, the algebra $\cS^0[\xi]$ endowed with the Hamiltonian gradation identifies to the graded algebra of complex functions on the graded manifold $T^*[2]M\times_M T[1]M$, the fiber variables $\xi$ and $p$ being considered of degree $1$ and $2$, see \cite{Roy02}. The symplectic form is then homogeneous and turns $\cS^0[\xi]$ into a graded Poisson algebra. This approach is well-suited to deal with classical mechanics since the momentum $\cJa_X$ (see \ref{applimomenta}), the natural and Darboux coordinates  $p_i$ and $\tp_i$ (see \ref{CoordDarboux}), are then all homogeneous, in $\cS^\d_{(1)}[\xi]$.

\begin{rmk}
The Hamiltonian gradation, that we introduce, is reminiscent of the gradation $\cS^\d_l[\xi]=\bigoplus_{k+\k=l}\cS^\d_{k,\k}[\xi]$, used by Getzler \cite{Get83} to prove the Atiyah-Singer index theorem.
\end{rmk}

\subsubsection{The module of spinor differential operators $\Dslm$}

The space of spinor densities of weight $\l$ is simply 
$
\sF^\l=\Ga(\sS)\otimes\cF^\l.
$
It admits a structure of $\conf(M,g)$-module, relying on the Lie derivative of spinor fields $\sL$, and given by 
$$
\sLl=\sL_X\otimes\Id+\Id\otimes\Ll,
$$
for all $X\in\conf(M,g)$. This enables us to define the module of differential operators acting on spinor densities.
\begin{defi}
The module $\Dslm$ is the space of differential operators $A:\sF^\l\rightarrow\sF^\mu$, endowed with the adjoint action of $\conf(M,g)$,
\begin{equation}\label{DefLD}
\LD A= \sLm A- A\sLl.
\end{equation}
\end{defi}
The natural filtration on $\Dslm$ is given by the order and the Hamiltonian filtration is defined along the same line as in Definition \ref{DefNHFiltre}. Forgetting the weights, it is given by
$$
\D_{(\frac{k}{2})}(M,\sS)=\mathrm{span}\bigcup_{2j+\k=k}\D_{j,\k}(M,\sS).
$$ 
The bracket is clearly compatible with the Hamiltonian filtration, i.e.\ $[\sD^{\l,\l}_{(\frac{k}{2})},\sD^{\l,\l}_{(\frac{k'}{2})}]\subset \sD^{\l,\l}_{(\frac{k+k'}{2}-1)}$, but not with the natural filtration since $[\g^i,\g^j]=-2g^{ij}$.
\begin{prop}\label{LtcL}
Let $\l\in\bbR$. The graded Poisson algebra associated to $\sD^{\l,\l}$ via its Hamiltonian filtration is isomorphic to the graded Poisson algebra $\cS^0[\xi]$. In particular they are isomorphic as $\conf(M,g)$-modules.
\end{prop}
\begin{proof}
The geometric quantization is well-defined on $\cS^0_{(1)}[\xi]$ and clearly extends uniquely as an algebra morphism to the graded algebra associated to $\D(M,\sS)$. Moreover, by Leibniz rule, the latter extension of geometric quantization still preserves the Poisson structures. Besides, the action of $X\in\conf(M,g)$ is given respectively by the Poisson bracket with the even comoment $\cJa_X$ and by the commutator with the Lie derivative of spinors $\sL_X$. Hence, the equality $\cQ_{\GQ}(\cJa_X)=(\hbar/\bi) \sL_X$ implies that both algebras are also isomorphic as $\conf(M,g)$-modules. 
\end{proof}

\subsection{The three $\conf(M,g)$-modules, in the conformally flat case}
Until the end of Section $5$, we suppose that $(M,g)$ is a conformally flat manifold.

\subsubsection{Comparison of the two modules of symbols}

Starting with conformal coordinates $(x^i)$ on $M$, we obtain natural coordinates $(x^i,p_i,\xi^i)$ and conformal Darboux coordinates $(x^i,\tp_i,\txi^i)$ (see Proposition \ref{PropConfDarboux}), providing the following local isomorphism: 
\begin{eqnarray}\label{Tenseur_Poly_Moments}
\ev_g:\cT^\d[\xi]&\stackrel{\text{\scriptsize{loc.}}}{\longrightarrow}& \cS^\d[\xi]\\ \nonumber
P^{i_1\dots i_k}_{j_1\ldots j_\k}(x)\, \xi^{j_1}\ldots \xi^{j_\k}  p_{i_1}\ldots p_{i_k} &\longmapsto&
|\vol_x|^{\frac{\k}{n}}P^{i_1\dots i_k}_{j_1\ldots j_\k}(x)\,  \txi^{j_1}\ldots \txi^{j_\k}  \tp_{i_1}\ldots \tp_{i_k}, 
\end{eqnarray} 
where $\vol_x=dx^1\wedge\ldots\wedge dx^n$. This local identification makes it easy to compare the module structures of $\cS^\d[\xi]$ and $\cT^\d[\xi]$.
\begin{prop}\label{PropLtLd}
Let $(x^i,\tp_i,\txi^i)$ be a conformal Darboux coordinate system  on $\cM$, $(\tilde{\partial}_i,\partial_{\tp_i},\partial_{\txi^i})$ the associated local basis and $\ev_g$ the isomorphism defined by (\ref{Tenseur_Poly_Moments}). The actions of $\conf(M,g)$ on the two modules $\cS^\d[\xi]$ and $\cT^\d[\xi]$ are related locally by
\begin{equation}\label{EqLtLd}
\medbox{\Ld=\ev_g \Lt \ev_g^{-1}-\frac{\hbar}{2\bi} \txi_k\txi^j(\partial_i\partial_j X^k)\partial_{\tp_i}.}
\end{equation}
Hence, $\cT^\d[\xi]$ is the graded $\conf(M,g)$-module associated to $\cS^\d[\xi]$ for the natural filtration.
\end{prop}
\begin{proof}
The formula (\ref{Ld}) giving $\Ld$ together with the expression (\ref{tildeX}) of $\tilde{X}$ in a Darboux coordinates system lead to 
$$
\left(\Ld+\frac{\hbar}{2\bi} \txi_k\txi^j(\partial_i\partial_j X^k)\partial_{\tp_i} \right)
=X^i \tilde{\partial_i}+\half\left((\partial_j X^i)\tilde{\xi}^j\partial_{\txi^i}-(\partial_i X^j)\tilde{\xi}_j\partial_{\txi_i}\right)
-\tp_j(\partial_i X^j)\partial_{\tp_i}+\d \partial_iX^i.
$$
We pull-back this equation to $\cT^\d[\xi]$ using the local isomorphism $\ev_g$ and get
\begin{eqnarray*}
(\ev_g)^{-1}\left(\Ld+\frac{\hbar}{2\bi} \txi_k\txi^j(\partial_i\partial_j X^k)\partial_{\tp_i} \right)\ev_g
&=&
X^i \partial_i+\half\left((\partial_j X^i)\xi^j\partial_{\xi^i}-(\partial_i X^j)\xi_j\partial_{\xi_i}\right)\\
&&
+(\partial_iX^i)\frac{\Sigma}{n}-p_j(\partial_i X^j)\partial_{p_i}+\left(\d-\frac{\Sigma}{n}\right) \partial_iX^i.
\end{eqnarray*}
To prove (\ref{EqLtLd}), we have to consider the term $\Xi=\xi^i\partial_iX^j\partial_{\xi^j}$ in formula (\ref{Lt}) giving $\Lt$. It can be written as ${\Xi=\half\left((\partial_j X^i)\xi^j\partial_{\xi^i}+(\partial_i X^j)\xi_j\partial_{\xi_i}\right)+\half\left((\partial_j X^i)\xi^j\partial_{\xi^i}-(\partial_i X^j)\xi_j\partial_{\xi_i}\right)}$, the first term being of symmetric type and the second one of skew-symmetric type. Moreover, since $X\in\conf(M,g)$ and $g$ is conformally flat, we have $g_{jk}\partial_iX^k+g_{ik}\partial_jX^k=\frac{2}{n}(\partial_kX^k)g_{ij}$. The first term of $\Xi$ is thus equal to $(\partial_iX^i)\frac{\Sigma}{n}$ and the result follows.
\end{proof}

\subsubsection{Comparison of $\Dslm$ with its two modules of symbols}

We already know that $\cS^\d[\xi]$ is the graded $\conf(M,g)$-module associated to $\sD^{\l,\m}$, for $\l=\m$. The extension to any couple $(\l,\m)$ is trivial, and we deduce from Proposition \ref{PropLtLd} the following corollary.
 \begin{cor}\label{CorLtLD}
 As $\conf(M,g)$-module, the space of tensorial symbols $\cT^\d[\xi]$ is the graded module associated to the natural filtration on $\Dslm$, for $\d=\m-\l$. 
  \end{cor}
  We can go further in the comparison of $\Dslm$ with its spaces of symbols via a local section of the principal symbol maps for both the natural and Hamiltonian filtration. Namely, geometric quantization of the supercotangent bundle can be extended to $\Om_\bbC(M)$ via the Weyl quantization $\g:\txi^i\mapsto\tg^i$, defined in Corollary \ref{GradCliff}. Then, choosing a Darboux coordinate system $(x^i,\tp_i,\txi^i)$, a further extension is provided by the normal ordering, which establishes a local isomorphism between the vector spaces $\cS^\d[\xi]$ and $\Dslm$,
{\begin{eqnarray}\label{OrdreNormal}
\cN:\cS^\d[\xi] & \stackrel{\text{\scriptsize{loc.}}}{\longrightarrow} &\Dslm   \\ \nonumber
P^{i_1\ldots i_k}_{j_1\ldots j_\k}(x)\;\txi^{j_1}\ldots\txi^{j_\k}\,\tp_{i_1}\ldots\tp_{i_k} &\longmapsto&  P^{i_1\ldots i_k}_{j_1\ldots j_\k}(x) \frac{\tg^{j_1}}{\sqrt{2}}\ldots\frac{\tg^{j_\k}}{\sqrt{2}}\,\frac{\hbar}{\bi}\partial_{i_1}\ldots\frac{\hbar}{\bi}\partial_{i_k}.
\end{eqnarray}}
\begin{prop}\label{LXcLX}
Let $(x^i,\tp_i,\txi^i)$ be a conformal Darboux coordinate system on $\cM$ and $\cN$ the associated normal ordering. For all $X\in \conf(M,g)$, we have, if $\d=\mu-\l$, 
\begin{equation}\label{cL-L} 
\medbox{\cN^{-1}\LD\cN=\Ld+\frac{\hbar}{2\bi}(\partial_j\partial_kX^i)\left(-\tp_i\partial_{\tp_j}+\half\chi^j_i\right)\partial_{\tp_k}
-\frac{\hbar}{\bi}\l\partial_j(\partial_iX^i)\partial_{\tp_j},}
\end{equation}
where $\chi^j_i=\txi^j\partial_{\txi^i}-\txi_i\partial_{\txi_j}+\half\partial_{\txi_j}\partial_{\txi^i}$.
\end{prop}
\begin{proof}
Let $P=P^{j_1\cdots j_k}_{i_1\cdots i_\k}(x)\,\frac{\tg^{i_1}}{\sqrt{2}}\cdots\frac{\tg^{i_\k}}{\sqrt{2}}\,\frac{\hbar}{\bi}\partial_{j_1}\cdots\frac{\hbar}{\bi}\partial_{j_k} \in 
\sD^{\l,\mu}$ and $X\in \conf(M,g)$.
By definition of the Lie derivative $\LD$ on $\Dslm$, we obtain
$$
\LD P = [\sL_X,P] + \d(\partial_iX^i)P-\l [P,(\partial_iX^i)].
$$
 Since $X\in\conf(M,g)$ is of degree at most two, the last term leads to $-\l [P,(\partial_iX^i)]=-\frac{\hbar}{\bi}\l\partial_j(\partial_iX^i)\partial_{\tp_j}$. The first term can be decomposed via the Leibniz rule, considering $P$ as the product of two terms,
\begin{eqnarray*}
[\sL_X,P] &=&
\left[\sL_X,P^{j_1\cdots j_k}_{i_1\cdots i_\k}(x)\frac{\tg^{i_1}}{\sqrt{2}}\cdots\frac{\tg^{i_\k}}{\sqrt{2}}\right]\frac{\hbar}{\bi}\partial_{j_1}\cdots\frac{\hbar}{\bi}\partial_{j_k}\\[6pt]
&&+ P^{j_1\cdots j_k}_{i_1\cdots i_\k}(x)
\frac{\tg^{i_1}}{\sqrt{2}}\cdots\frac{\tg^{i_\k}}{\sqrt{2}}
\left[\sL_X,\frac{\hbar}{\bi}\partial_{j_1}\cdots\frac{\hbar}{\bi}\partial_{j_k}\right].
\end{eqnarray*}
Let $P_0\in\Om_\bbC(M)$. Theorem \ref{thmSpinorL} together with the definition of $\cN$ prove that $\sL_X=\frac{\bi}{\hbar}\cN(\cJa_X)$; we hence get $[\sL_X,\cN(P_0)]=\frac{\bi}{\hbar}[\cN(\cJa_X),\cN(P_0)]$. Moreover, the normal ordering $\cN$ coincides with the Weyl quantization on $\Om_\bbC(M)$ and satisfies then Equation (\ref{QWIsoLie}). Thereby, we have the equalities $[\sL_X,\cN(P_0)]=\cN(\{\cJa_X,P_0\})=\cN(\tilde{X} P_0)$. Denoting by $P_0^{j_1\dots j_k}$ the symbol $P_{i_1 \ldots i_\k}^{j_1\dots j_k}(x)\,\txi^{i_1}\ldots\txi^{i_\k}$, we end up with 
$$
\cN^{-1}\left([\sL_X,P]+\d(\partial_iX^i)P\right) = \Ld(P_0^{j_1\dots j_k})\tp_{j_1}\ldots\tp_{j_k}
+ \cN^{-1}\left(
\cN(P_0^{j_1\dots j_k})\Big[\sL_X,\frac{\hbar}{\bi}\partial_{j_1}\cdots\frac{\hbar}{\bi}\partial_{j_k}\Big]\right).
$$
Since $X\in\conf(M,g)$ is of degree at most two, the last term decomposes as
\begin{eqnarray*}
\cN(P_0^{j_1\dots j_k})
\Big[\sL_X,\frac{\hbar}{\bi}\partial_{j_1}\cdots\frac{\hbar}{\bi}\partial_{j_k}\Big]
&=&\cN(P_0^{j_1\dots j_k})\frac{\hbar}{\bi}{[\sL_X,\partial_i]}\cN(\partial_{\tp_i}(\tp_{j_1}\ldots\tp_{j_k}))\\ 
&&+\cN(P_0^{j_1\dots j_k})\left(\frac{\hbar}{\bi}\right)^2{[[\sL_X,\partial_i],\partial_j]}\cN(\partial_{\tp_i}\partial_{\tp_j}(\tp_{j_1}\ldots\tp_{j_k})).
\end{eqnarray*}
The  coordinates $(x^i)$ are conformal, thus the identity $\partial_i=\sL_{\partial_i}$ holds. As $[[X,\partial_i]\partial_j]$ is constant, it is equal to $\partial_j(\partial_iX^k)\partial_k$ and the term on the second line is determined.
 Besides, the formula (\ref{sLX}), which gives the spinor Lie derivative, allows us to prove that 
$$
{[\sL_X,\partial_i]}= -(\partial_iX^j)\partial_j
+A_i,
$$
where $A_i=\frac{1}{8}\partial_i\left[\partial_k (X^j)\g^k\g_j-\partial_k (X^j)\g_j\g^k\right]$. Let $\tilde{A}_i$ be the operator on $\Cinfty(\cM)$ such that 
$
[\cN(P_0^{j_1\dots j_k}),A_i]=\cN(\tilde{A}_i\cdot P_0^{j_1\dots j_k}).
$
Denoting by $\chi^k_j=\txi^k\partial_{\txi^j}-\txi_j\partial_{\txi_k}+\half\partial_{\txi_k}\partial_{\txi^j}$, we obtain 
$$
\tilde{A}_i=\half\partial_i\partial_kX^j(2\txi^k\txi_j+\chi^k_j),
$$
thanks to the equality 
$
\gamma^{i_1}\cdots\gamma^{i_\k}(\gamma^{k}\gamma^{j})
=(\gamma^{k}\gamma^{j})\gamma^{i_1}\cdots\gamma^{i_\k}
-\left[\gamma^k\gamma^j,\gamma^{i_1}\cdots\gamma^{i_\k}\right]
$ and the Proposition~\ref{gLie}.
Combining the above equalities and the expression (\ref{Ld}) of $\Ld$ leads to the required formula~(\ref{cL-L}).
\end{proof}

\subsubsection{The explicit actions of $\conf(M,g)$ on the $3$ modules}

We work locally with a fixed conformal coordinate system $(x^i)$ of $M$, and the induced conformal Darboux coordinates $(x^i,\tp_i,\txi^i)$ on $\cM$. Its associated local basis is denoted $(\tilde{\partial}_i,\partial_{\tp_i},\partial_{\txi^i})$. We write the actions of every generator of $\conf(M,g)$, as defined in (\ref{GenConf}), in terms of those conformal Darboux coordinates, pulling-back the actions on $\Dslm$ to $\cS^\d[\xi]$ via $\cN$ (see \ref{OrdreNormal}), and pushing-forward those on $\cT^\d[\xi]$ to $\cS^\d[\xi]$ via $\ev_g$ (see \ref{Tenseur_Poly_Moments}). 

From Propositions \ref{PropLtLd} and \ref{LtcL} we deduce that the actions of $X\in\ce(p,q)$ coincide on each of the modules $\cT^\d[\xi]$, $\cS^\d[\xi]$ and $\Dslm$, modulo the isomorphisms $\ev_g$ and $\cN$, namely
\begin{equation}\label{Lcepq}
\ev_g\Lt\ev_g^{-1}=\Ld=\cN\LD\cN^{-1}.
\end{equation}
Using the explicit expression (\ref{Lt}) of $\Lt$, we deduce the action of the generators of the affine conformal transformations $\ce(p,q)$ on $\cS^\d[\xi]$. For $i,j=1,\ldots,n$, we find
\begin{eqnarray}\nonumber
L_{X_i}^\d&=&\tilde{\partial_i},\\ \label{LdX0}
L_{X_{ij}}^\d&=& x_i \tilde{\partial_j}- x_j \tilde{\partial_i}+\tp_i\partial_{\tp^j}- \tp_j\partial_{\tp^i}+\txi_i\partial_{\txi^j}-\txi_j\partial_{\txi^i},\\  \nonumber
L_{X_0}^\d&=& x^i \tilde{\partial_i}-\tp_i\partial_{\tp_i}+\d n.
\end{eqnarray}

We still have to compute the explicit actions of the inversions $\bar{X}_i$ on the three modules $\cT^\d[\xi]$, $\cS^\d[\xi]$ and $\Dslm$. From the general formulas (\ref{Lt}) and (\ref{Ld}), we deduce the action of inversions on $\cT^\d[\xi]$, modulo $\ev_g$, and $\cS^\d[\xi]$, viz.
\begin{eqnarray}\label{LtXi}   
\ev_g\bbL^\d_{\bar{X}_i}\ev_g^{-1}&=&(x_jx^j\tilde{\partial}_i - 2x_ix^j\tilde{\partial}_j) +(-2p_ix_j\partial_{\tp_j}+2x_ip_j\partial_{\tp_j}+2p_k x^k\partial_{\tp^i})\\ \nonumber
&&
+2x_j\xi^j\partial_{\txi^i}-2\xi_i x^k\partial_{\txi^k}
-2n\d x_i,\\[6pt] \label{LdXi}
L_{\bar{X}_i}^\d&=&\ev_g \bbL^\d_{\bar{X}_i} (\ev_g)^{-1}-2\frac{\hbar}{\bi} \txi_i\txi^j\partial_{\tp_j},
\end{eqnarray}
for all $i=1,\ldots,n$. The general formula (\ref{cL-L}) can be particularized to the inversion generators $\bar{X}_i$, for $i=1,\ldots,n$, and leads to the action of the inversions on $\Dslm$, namely 
\begin{equation}\label{cL-LXi}
\cN^{-1}\,\ccL^{\l,\m}_{\bar{X}_i}\,\cN= L_{\bar{X}_i}^\d +\frac{\hbar}{\bi}(-\tp_i\partial_{\tp^j}\partial_{\tp_j}+2\tp_j\partial_{\tp_j}\partial_{\tp^i})
+\frac{\hbar}{\bi}\chi^j_i\partial_{\tp_j}
+2\frac{\hbar}{\bi}n\l\partial_{\tp_i}.
\end{equation}  

\subsection{Conformally invariant elements of the modules $\cT^\d[\xi]$, $\cS^\d[\xi]$ and $\Dslm$}
 
From now on, we suppose that $(M,g)$ is an oriented and conformally flat manifold, so $\vol_g$ is a globally defined volume form on $M$.
We will classify the conformally invariant elements of each of the three module families $(\cT^\d[\xi])_{\d\in\bbR}$, $(\cS^\d[\xi])_{\d\in\bbR}$ and $(\Dslm)_{\l,\m\in\bbR}$, namely those elements in the kernel of the action of $\conf(M,g)$. We follow the strategy adopted in \cite{ORe03}, concerning the modules of scalar differential operators and of their symbols. It relies on the determination of Euclidean invariants by Weyl's theory of invariants \cite{Wey97} and on the explicit actions of dilation and inversions on each of the three families of modules.

We first classify the isometry invariants, in terms of conformal Darboux coordinates $(x^i,\tp_i,\txi^i)$. 
Resorting to Equation (\ref{Lcepq}), we see that the structures of $\e(p,q)$-modules $\cT^\d[\xi]$, $\cS^\d[\xi]$ and $\Dslm$ are isomorphic and do not depend on the weights $\l$ and $\d=\m-\l$. Hence, their isometry invariants coincide via the identification maps $\ev_g$ and $\cN$, and we will explicitly give those of the module $\cS^\d[\xi]$. For $\d=0$, they form an algebra which is the commutator $\e(p,q)^!$ of the action of isometries in $\cS^0[\xi]$. 
\begin{prop}\label{PropRDeltaChi}
Let $n=p+q\geq 2$. The algebra $\e(p,q)^!$ of local isometry invariants of the module $\cS^0[\xi]$ is generated by
$$
 \tilde{\chi}=(\vol_x)_{j_1 \ldots j_n}\txi^{j_1}\ldots \txi^{j_n},\quad \tilde{\Delta}=\txi^i \tp_i,  \quad \tilde{\Delta}\star\tilde{\chi} =(\vol_x)_{j_1 \ldots j_n}\tp^{j_1}\txi^{j_2}\ldots \txi^{j_n}  \;\; \text{and}\; \;\tilde{R}=\eta^{ij}\tp_i\tp_j,
$$
with $\eta$ the flat metric of signature $(p,q)$ and $\vol_x$ the associated volume form. The local isometry invariants of $\cS^\d[\xi]$ form the space $|\vol_x|^\d\cdot\e(p,q)^!$.
\end{prop}  
\begin{proof}
Since the local action of $\e(p,q)$ on $\cS^0[\xi]$ is of tensorial type, we can apply a classical theorem of Weyl \cite{Wey97}, which states that all $\ro(p,q)$-invariants are constructed from the metric and the volume form, i.e$.$ from $\eta$ and $\vol_x$. Moreover, the isometry invariants must be independent of the coordinates $(x^i)$ since they are invariant under translations. We easily deduce that $\tilde{R}$, $\tilde{\Delta}$, $\tilde{\chi}$ and $\tilde{\Delta}\star\tilde{\chi}$ generate $\e(p,q)^!$. As $|\vol_x|$ is invariant under the action of $\e(p,q)$, we deduce the case of a general weight $\d$. 
\end{proof}
Let us remark that $\tilde{\Delta}\star\tilde{\chi}$ is the Moyal product of $\tilde{\Delta}$ and $\tilde{\chi}$, i.e$.$ $\tilde{\Delta}\star\tilde{\chi}=-\frac{\hbar}{\bi}\{\tilde{\Delta},\tilde{\chi}\}$. The Proposition \ref{PropRDeltaChi} may then be rephrased as follows: the algebra $\e(p,q)^!$ is vectorially generated by $\tilde{\Delta}^a\star\tilde{\chi}^b\,\tilde{R}^s$, where $a,b=0,1$ and $s\in\bbN$.
In order to be invariant under the action (\ref{LdX0}) of dilation, the symbols generating $\e(p,q)^!$ must have a weight: $\d=\frac{2}{n}$ for $\tilde{R}$, $\d=\frac{1}{n}$ for $\tilde{\Delta}$ and $\tilde{\Delta}\star\tilde{\chi}$, and $\d=0$ for $\tilde{\chi}$. They admit globally defined analogues.

\begin{defi}\label{GlobalRDeltaChi}
Let $(x^i,p_i,\xi^i)$ be natural coordinates on the supercotangent bundle of $(M,g)$, and $\vol_g$ be the volume form induced by $g$. We define the following global symbols:
$$
\begin{array}{c}
\chi=(\vol_g)_{j_1\dots j_n}\xi^{j_1}\ldots\xi^{j_n}\in\cS^0[\xi],\quad \Delta =|\vol_g|^{\frac{1}{n}} \left(p_i\xi^i\right)\in\cS^\frac{1}{n}[\xi], \\[6pt]
\Delta\star\chi = |\vol_g|^{\frac{1}{n}}\left(g^{ij_1}\,(\vol_g)_{j_1\dots j_n}p_i\xi^{j_2}\ldots\xi^{j_n}\right)\in\cS^\frac{1}{n}[\xi]\quad\textit{and}\quad R=|\vol_g|^{\frac{2}{n}}\left(g^{ij}p_ip_j\right)\in\cS^\frac{2}{n}[\xi].
\end{array}
$$
\end{defi}
In general, the conformally invariant elements on a conformally flat manifold $(M,g)$ are globally defined, and this holds clearly true for the elements of the modules of tensors $\cT^\d[\xi]$ and of operators $\Dslm$. Besides, we deduce from Proposition \ref{PropConfDarboux} that all expressions in conformal Darboux coordinates, which are invariant under the Hamiltonian action of $\conf(M,g)$, are globally defined on $\cM$, and that stands in particular for elements of $\cS^\d[\xi]$.  
Therefore, we will use Definition \ref{GlobalRDeltaChi} to give a global expression for the conformal invariants of the module families $(\cT^\d[\xi])_{\d\in\bbR}$ and $(\cS^\d[\xi])_{\d\in\bbR}$. Nevertheless, we will only provide local expressions, via normal ordering, for the conformally invariant differential operators. To obtain their global expressions is a difficult matter, even in the case of scalar differential operators \cite{Pan08,GPe03}.
\begin{thm}\label{ThmClassifOpDiffSpin}
The conformal invariants of the module family $(\cT^\d[\xi])_{\d\in\bbR}$ are given by,
\begin{equation}\label{InvConfcT}
|\vol_g|^{-\frac{a}{n}-b}\,\Delta^a\star\chi^b\, R^s \in\cT^{\frac{2s+a}{n}}[\xi],
\end{equation}
where $a,b=0,1$ and $s\in\bbN$. Those of the module family $(\cS^\d[\xi])_{\d\in\bbR}$ read
\begin{equation}\label{InvConfcS}
 \Delta^a\star\chi^b\, R^s \in\cS^{\frac{2s+a}{n}}[\xi],
\end{equation}
where $s\in\bbN$ and  $a,b=0,1$ with $a+b\neq 0$.
The conformal invariants of the module family $(\Dslm)_{\l,\m\in\bbR}$ retain, via the normal ordering, the local expression
\begin{equation}\label{InvConfDlm}
\cN(\chi)\in\sD^{\l,\l},\qquad \cN(\Delta),\,\cN(\Delta\star\chi)\in\sD^{\frac{n-1}{2n},\frac{n+1}{2n}},\qquad  \cN(\Delta\, R^s) \in\sD^{\frac{n-2s-1}{2n},\frac{n+2s+1}{2n}},
\end{equation}
where $s\in\bbN$ and $\l\in\bbR$. 
\end{thm}
\begin{proof}
Redefining $\tilde{\Delta}=|\vol_x|^\frac{1}{n}\tp_i\txi^i$ and $\tilde{R}=|\vol_x|^\frac{2}{n}\eta^{ij}\tp_i\tp_j$,  
the local invariants under the Lie algebra $\ce(p,q)$ are, modulo $(\ev_g)^{-1}$ and $\cN$, of the form $\tilde{\Delta}^a\star\tilde{\chi}^b\,\tilde{R}^s $, with $a,b=0,1$ and $s\in\bbN$. Using the expression (\ref{tptxi}) of conformal Darboux coordinates, we notice that $\chi=\tilde{\chi}=|\vol_g|\ev_g^{-1}(\tilde{\chi})$, $\Delta=\tilde{\Delta}=|\vol_g|^{\frac{1}{n}}\ev_g^{-1}(\tilde{\Delta})$, but $R=\ev_g^{-1}(\tilde{R})$ and $R=\tilde{R}-\frac{\hbar}{\bi}\Delta \frac{\xi^j\partial_jF}{F}$, where $F$ is the conformal factor: $g_{ij}=F\eta_{ij}$. Now we obtain the classification of conformal invariants of, respectively, $\cT^\d[\xi]$, $\cS^\d[\xi]$ and $\Dslm$.
\begin{enumerate}
\item The action of conformal inversions on the module $\cT^\d[\xi]$ is given in (\ref{LtXi}), and vanishes on the local $\ce(p,q)$-invariants $\ev_g^{-1}(\tilde{\Delta}^a\star\tilde{\chi}^b\,\tilde{R}^s)$. Hence, the latter are conformally invariant in $(\cT^\d[\xi])_{\d\in\bbR}$, and they take the global expressions announced in (\ref{InvConfcT}).
\item For the module $\cS^\d[\xi]$, the action of inversions is given by (\ref{LdXi}), and then, for $\d=\frac{2s+a}{n}$,
$$
L^\d_{\bar{X}_i}\left(\tilde{\Delta}^a\star\tilde{\chi}^b\tilde{R}^s\right) =-2\frac{\hbar}{\bi}\txi_i \tilde{\Delta}^a\star\tilde{\chi}^b\,[\txi^i\partial_{\tp_i},\tilde{R}^s],
$$
since $[\txi^i\partial_{\tp_i},\tilde{\Delta}]=[\txi^i\partial_{\tp_i},\tilde{\chi}]=0$. Now $[\txi^i\partial_{\tp_i},\tilde{R}^s]=2s\tilde{R}^{s-1}\tilde{\Delta}$, implying that the symbol $\tilde{\Delta}^a\star\tilde{\chi}^b\,\tilde{R}^s$ is conformally invariant if and only if $a+b\neq 0$. In these cases, we easily check the equality $\tilde{\Delta}^a\star\tilde{\chi}^b\,\tilde{R}^s=\Delta^a\star\chi^b\,R^s$, hence the result (\ref{InvConfcS}).

\item At last, the action of inversions on the module $\Dslm$ is given by (\ref{cL-LXi}), and we can evaluate it on the similarity invariants $\tilde{\Delta}^a\star\tilde{\chi}^b\,\tilde{R}^s$, modulo normal ordering.
Firstly, the symbol $\tilde{\chi}$, of weight $\d=0$, clearly vanishes under this action if $\l=\m$. Secondly, we get, for  $\d=\m-\l=\frac{2s}{n}$, 
$$
\ccL^{\l,\m}_{\bar{X}_i}\tilde{R}^s =2s\frac{\hbar}{\bi}\left[(2n\l+2s-n)\tp_i-2\txi_i\tilde{\Delta}\right]\tilde{R}^{s-1}.
$$
Thus, $\tilde{R}^s$ is not conformally invariant if $s\neq 0$. Similarly, the action of  $\ccL^{\l,\m}_{\bar{X}_i}$ on $\tilde{\chi} \tilde{R}^s$ is 
$$
\ccL^{\l,\m}_{\bar{X}_i}\tilde{\chi} \tilde{R}^s =2s\frac{\hbar}{\bi}\left[(2n\l+2s-n)\tp_i-\partial_{\txi^i}\tilde{\Delta}\star\tilde{\chi}\right]\tilde{R}^{s-1},
$$
which is a nonvanishing expression if $s\neq 0$.
We still have to evaluate this action on $\tilde{\Delta}\star\tilde{\chi}^b \tilde{R}^s$, for $b=0,1$. On the one hand, we obtain, for $\mu-\l=\frac{2s+1}{n}$
$$
\ccL^{\l,\m}_{\bar{X}_i}\tilde{\Delta} \tilde{R}^s =\frac{\hbar}{\bi}\left(2s+1-n+2n\l \right)\left(\txi_i\tilde{R}^s+2s\tp_i\tilde{\Delta} \tilde{R}^{s-1}\right) ,
$$
which vanishes if and only if $\l=\frac{n-2s-1}{2n}$, and on the other hand
$$
\ccL^{\l,\m}_{\bar{X}_i}\tilde{\Delta}\star\tilde{\chi} \tilde{R}^s =2s\frac{\hbar}{\bi}\left(2s-n+2n\l \right)\tp_i\tilde{\Delta}\star\tilde{\chi} \tilde{R}^{s-1}+\frac{\hbar}{\bi}\left(4s+1-n+2n\l \right)\partial_{\txi_i}\tilde{\chi} \tilde{R}^s ,
$$
which vanishes if and only if $s=0$ and $\l=\frac{n-1}{2n}$. The result (\ref{InvConfDlm}) follows.
\end{enumerate}
The proof of Theorem \ref{ThmClassifOpDiffSpin} is complete.
\end{proof}
\begin{rmk}
All the conformal invariants of the module families $(\cT^\d[\xi])_{\d\in\bbR}$ and $(\Dslm)_{\l,\m\in\bbR}$ are invariant under a conformal rescaling of the metric $g\mapsto F^2g$. Such a transformation affects indeed $\g^i=\cN(\xi^i)$ and $|\vol_g|^{-\frac{1}{n}}\xi^i$  by a coefficient $F^{-1}$. In contradistinction, no conformal invariant of the module family $(\cS^\d[\xi])_{\d\in\bbR}$ admits such an invariance. Recalling that $\cS^\d[\xi]=\Ga(\cS TM\otimes\Lambda T^*M)\otimes\cF^\d$, this is a consequence of the non homogeneity of the bundle $\cS TM\otimes\Lambda T^*M$ for the Hamiltonian action of $\conf(M,g)$, as shown by the explicit action of the infinitesimal inversions (\ref{LdXi}). 
\end{rmk}

As the conformally invariant differential operators of degree $1$ or less are well-known the following theorem is trivially deduced from the last one.
\begin{thm}
The conformally invariant differential operators of order $p$ acting on weighted spinors are, for $s\in\bbN^*$,
\begin{enumerate}
\item if $p=0$, the chirality: $(\vol_g)_{i_1\cdots i_n}\g^{i_1}\ldots\g^{i_n}\in\sD^{\l,\l}$,
\item if $p=1$, the Dirac operator: $\g^i\nabla_i\in\sD^{\frac{n-1}{2n},\frac{n+1}{2n}}$,
or the twisted Dirac operator:\\ $g^{ij_1}(\vol_g)_{j_1\ldots j_n}\g^{j_2}\ldots \g^{j_n}\nabla_{i}\in\sD^{\frac{n-1}{2n},\frac{n+1}{2n}}$,
\item if $p=2s$, no operator,
\item if $p=2s+1$, the operator in $\sD^{\frac{n-2s-1}{2n},\frac{n+2s+1}{2n}}$ given locally by $\cN(\Delta\, R^s)$.
\end{enumerate}
\end{thm}
In the latter theorem, we recover one of the two families of operators in $\Dslm$, shown to be invariant under rescalings of the metric by Holland and Sparling \cite{HSp01}, namely the one of the conformal odd powers of the Dirac operators. Note that global explicit expressions have been found for these operators over spheres in \cite{ESo10}. 
The two additional invariants that we get depend on the orientation of $M$. In accordance with Branson's work on conformally invariant operators of second order \cite{Bra98}, we get no invariant differential operators of even order.

\section{Outlook}
We present open questions on the constructions of the spinor bundle and of the Lie derivatives of spinors, as they are performed in this paper. Afterwards, we put forward the conformally equivariant quantization of the supercotangent bundles as a natural continuation of the present work. This is the purpose of a paper in preparation, and is already partly developed in~\cite{Mic09}. 

The supercotangent bundle $\cM$ of a pseudo-Riemannian manifold $(M,g)$ possesses a canonical symplectic form $\om$, as proved by Rothstein \cite{Rot90}. Upon topological conditions on~$M$, we have been able to perform the geometric quantization of $(\cM,\om)$, and, thereby, to construct the spinor bundle of $M$ as well as the covariant derivative and the Lie derivative of spinors. These three objects exist as soon as $M$ admits a spin structure. We have proved that we need stronger hypothesis on $(M,g)$ to apply geometric quantization to $(\cM,\om)$, at least for $g$ a Riemannian metric. It would be nice to weaken those hypothesis in order that they precisely coincide with the existence of a spin structure on $M$. This would probably require a generalization of the notion of polarization, as, for example, that of higher polarization proposed in the framework of Group Approach to Quantization \cite{AGM98}.

Finding a Hamiltonian lift of the vector fields of $M$ to its supercotangent bundle has proved to be non trivial. That is not surprising, by means of geometric quantization, such a lift corresponds indeed to a Lie derivative of spinors. In particular, the lift that we have chosen is quantized as the Lie derivative of spinors of Kosmann, defined on the Lie subalgebra $\conf(M,g)$ of $\Vect(M)$. Do there exist other Hamiltonian lifts, i.e$.$ other Lie derivatives of spinors than Kosmann's? On which Lie subalgebras of vector fields are they defined? What geometrical object plays the role of the odd $1$-form $\b$?

We have provided $\conf(M,g)$-module structures on each of the filtered spaces of Hamiltonian symbols $\cS^\d[\xi]$ and of spinor differential operators $\Dslm$, as well as on the associated graded module of tensorial symbols $\cT^\d[\xi]$. This extends the scalar case, for which there is a unique module of symbols $\cS^\d=\Pol(T^*M)\otimes\cF^\d$ associated to the module of scalar differential operators $\Dlm$. Duval, Lecomte and Ovsienko have shown that, in the conformally flat case, the $\conf(M,g)$-modules $\cS^\d$ and $\Dlm$ are isomorphic if $\d=\m-\l$, and the isomorphism is unique if we require the preservation of the principal symbol \cite{DLO99}. Such an isomorphism is called a conformally equivariant quantization; its inverse is a symbol map. More precisely, the existence and uniqueness of a conformally equivariant quantization was proved for generic values of the weights $\l,\m$, in the scalar case \cite{DLO99}. The exceptional values are called resonances. Naturally, we can ask for a generalization to the spin case, which, in fact, holds true. 
\begin{thm}\cite{Mic09}
Let $(M,g)$ be a conformally flat manifold and $\d=\m-\l\in\bbR$. There exists (for generic $\l$, $\m$) a unique conformally equivariant quantization $\cQlm:\cS^\d[\xi]\rightarrow\Dslm$, i.e$.$ a unique isomorphism of $\conf(M,g)$-module, preserving the principal symbol.
\end{thm}
We have also obtained a similar theorem in \cite{Mic09}, on the (generic) existence and uniqueness of a conformally equivariant superization $\rST^\d:\cT^\d[\xi]\rightarrow\cS^\d[\xi]$, whose name is due to the inclusion of modules, $\cS^\d\subset\cT^\d[\xi]$.

Straightforwardly, the conformal invariants of the three modules $\cT^\d[\xi]$, $\cS^\d[\xi]$ and $\Dslm$, correspond to each other via the conformally equivariant superization and quantization as soon as they exist. This is the case for the three conformal invariants of lower order \cite{Mic09}, in accordance with Theorem \ref{ThmClassifOpDiffSpin}. On the contrary, the same Theorem \ref{ThmClassifOpDiffSpin} proves that the invariant $R\in\cT^{\frac{2}{n}}[\xi]$ has no equivalent in $\cS^\frac{2}{n}[\xi]$, implying the non-existence of $\rST^\frac{2}{n}$.
A full investigation of this correspondence between the resonances of equivariant quantization and the existence of conformal invariants has been performed in \cite{Mic11a}. Besides, it would be interesting to obtain a geometric expression for the conformal third power of the Dirac operator that we get in our classification. In particular, it could lead to a new type of $Q$-curvature.

 \subsection*{Acknowledgements}
 It is a pleasure to acknowledge Christian Duval for his essential guidance in our investigation of geometric and conformally equivariant quantizations of $(\cM,d\a)$, and invaluable discussions. Special thanks are due to Valentin Ovsienko for his constant interest in this work.

\bibliographystyle{plain}
\bibliography{Biblio}

\begin{thebibliography}{10}

\bibitem{AGM98}
V.~Aldaya, J.~Guerrero, and G.~Marmo.
\newblock Quantization on a {L}ie group: higher-order polarizations.
\newblock In {\em Symmetries in science, {X} ({B}regenz, 1997)}, pages 1--36.
  Plenum, New York, 1998.

\bibitem{BCL77}
A.~{Barducci}, R.~{Casalbuoni}, and L.~{Lusanna}.
\newblock Classical spinning particles interacting with external gravitational
  fields.
\newblock {\em Nuclear Physics B}, 124:521--538, June 1977.

\bibitem{BMa77}
F.~A. {Berezin} and M.~S. {Marinov}.
\newblock Particle spin dynamics as the grassmann variant of classical
  mechanics.
\newblock {\em Annals of Physics}, 104:336--362, April 1977.

\bibitem{BGV92}
N.~Berline, E.~Getzler, and M.~Vergne.
\newblock {\em Heat kernels and {D}irac operators}.
\newblock Grundlehren Text Editions. Springer-Verlag, Berlin, 2004.
\newblock Corrected reprint of the 1992 original.

\bibitem{Bla72}
R.~J. Blattner.
\newblock Quantization and representation theory.
\newblock In {\em Harmonic analysis on homogeneous spaces ({P}roc. {S}ympos.
  {P}ure {M}ath., {V}ol. {XXVI}, {W}illiams {C}oll., {W}illiamstown, {M}ass.,
  1972)}, pages 147--165. Amer. Math. Soc., Providence, R. I., 1973.

\bibitem{Bor00}
M.~Bordemann.
\newblock The deformation quantization of certain super-{P}oisson brackets and
  {BRST} cohomology.
\newblock In {\em Conf{\'e}rence {M}osh{\'e} {F}lato 1999, {V}ol. {II}
  ({D}ijon)}, volume~22 of {\em Math. Phys. Stud.}, pages 45--68. Kluwer Acad.
  Publ., Dordrecht, 2000.

\bibitem{BGa92}
J.-P. Bourguignon and P.~Gauduchon.
\newblock Spineurs, op{\'e}rateurs de {D}irac et variations de m{\'e}triques.
\newblock {\em Comm. Math. Phys.}, 144(3):581--599, 1992.

\bibitem{Bra95}
T.~P. Branson.
\newblock Sharp inequalities, the functional determinant, and the complementary
  series.
\newblock {\em Trans. Amer. Math. Soc.}, 347(10):3671--3742, 1995.

\bibitem{Bra98}
T.~P. Branson.
\newblock Second order conformal covariants.
\newblock {\em Proc. Amer. Math. Soc.}, 126(4):1031--1042, 1998.

\bibitem{CGS10}
A.~Cap, A.~R. Gover, and V.~Soucek.
\newblock Conformally invariant operators via curved casimirs: Examples.
\newblock {\em Pure Appl. Math. Q.}, 6(3):693--714, 2010.

\bibitem{DeW92}
B.~DeWitt.
\newblock {\em Supermanifolds}.
\newblock Cambridge Monographs on Mathematical Physics. Cambridge University
  Press, Cambridge, second edition, 1992.

\bibitem{DLO99}
C.~Duval, P.~B.~A. Lecomte, and V.~Yu. Ovsienko.
\newblock Conformally equivariant quantization: existence and uniqueness.
\newblock {\em Ann. Inst. Fourier (Grenoble)}, 49(6):1999--2029, 1999.

\bibitem{Eas05}
M.~G. Eastwood.
\newblock Higher symmetries of the {L}aplacian.
\newblock {\em Ann. of Math. (2)}, 161(3):1645--1665, 2005.

\bibitem{ELe08}
M.~G. Eastwood and T.~Leistner.
\newblock Higher symmetries of the square of the {L}aplacian.
\newblock In {\em Symmetries and overdetermined systems of partial differential
  equations}, volume 144 of {\em IMA Vol. Math. Appl.}, pages 319--338.
  Springer, New York, 2008.

\bibitem{ERi87}
M.~G. Eastwood and J.~W. Rice.
\newblock Conformally invariant differential operators on {M}inkowski space and
  their curved analogues.
\newblock {\em Comm. Math. Phys.}, 109(2):207--228, 1987.
\newblock Erratum Comm. Math. Phys., 144(1): 213, 1992.

\bibitem{ESo10}
D.~Eelbode and V.~Sou{\v{c}}ek.
\newblock Conformally invariant powers of the {D}irac operator in {C}lifford
  analysis.
\newblock {\em Math. Methods Appl. Sci.}, 33(13):1558--1570, 2010.

\bibitem{ENi96}
A.~M. {El Gradechi} and L.~M. Nieto.
\newblock Supercoherent states, super-{K}{\"a}hler geometry and geometric
  quantization.
\newblock {\em Comm. Math. Phys.}, 175(3):521--563, 1996.

\bibitem{FLl00}
S.~Ferrara and M.~A. Lled{\'o}.
\newblock Some aspects of deformations of supersymmetric field theories.
\newblock {\em J. High Energy Phys.}, (5):Paper 8, 2000.

\bibitem{Get83}
E.~Getzler.
\newblock Pseudodifferential operators on supermanifolds and the
  {A}tiyah-{S}inger index theorem.
\newblock {\em Comm. Math. Phys.}, 92(2):163--178, 1983.

\bibitem{GMa05}
M.~Godina and P.~Matteucci.
\newblock The {L}ie derivative of spinor fields: theory and applications.
\newblock {\em Int. J. Geom. Methods Mod. Phys.}, 2(2):159--188, 2005.

\bibitem{GPe03}
A.~R. Gover and L.~J. Peterson.
\newblock Conformally invariant powers of the {L}aplacian, {$Q$}-curvature, and
  tractor calculus.
\newblock {\em Comm. Math. Phys.}, 235(2):339--378, 2003.

\bibitem{GSi09}
A.~R. Gover and J.~Silhan.
\newblock Higher symmetries of the conformal powers of the {L}aplacian on
  conformally flat manifolds.
\newblock {\em J. Math Phys.}, 53(3):26 pp., 2012.

\bibitem{GJMS92}
C.~R. Graham, R.~Jenne, L.~J. Mason, and G.~A.~J. Sparling.
\newblock Conformally invariant powers of the {L}aplacian. {I}. {E}xistence.
\newblock {\em J. London Math. Soc. (2)}, 46(3):557--565, 1992.

\bibitem{Hit74}
N.~Hitchin.
\newblock Harmonic spinors.
\newblock {\em Advances in Math.}, 14:1--55, 1974.

\bibitem{Khu91}
O.~M. Khudaverdian.
\newblock Geometry of superspace with even and odd brackets.
\newblock {\em J. Math. Phys.}, 32(7):1934--1937, 1991.

\bibitem{Kos72}
Y.~Kosmann.
\newblock D{\'e}riv{\'e}es de {L}ie des spineurs.
\newblock {\em Ann. Mat. Pura Appl. (4)}, 91:317--395, 1972.

\bibitem{Kos70}
B.~Kostant.
\newblock Quantization and unitary representations. {I}. {P}requantization.
\newblock In {\em Lectures in modern analysis and applications, {III}}, pages
  87--208. Lecture Notes in Math., Vol. 170. Springer, Berlin, 1970.

\bibitem{Kos74}
B.~Kostant.
\newblock Symplectic spinors.
\newblock In {\em Symposia {M}athematica, {V}ol. {XIV} ({C}onvegno di
  {G}eometria {S}implettica e {F}isica {M}atematica, {INDAM}, {R}ome, 1973)},
  pages 139--152. Academic Press, London, 1974.

\bibitem{Kos77}
B.~Kostant.
\newblock Graded manifolds, graded {L}ie theory, and prequantization.
\newblock In {\em Differential geometrical methods in mathematical physics
  ({P}roc. {S}ympos., {U}niv. {B}onn, {B}onn, 1975)}, pages 177--306. Lecture
  Notes in Math., Vol. 570. Springer, Berlin, 1977.

\bibitem{LMi89}
H.~B. Lawson and M.-L. Michelsohn.
\newblock {\em Spin geometry}, volume~38 of {\em Princeton Mathematical
  Series}.
\newblock Princeton University Press, Princeton, NJ, 1989.

\bibitem{Lei77}
D.~A. Le{\u\i}tes.
\newblock New {L}ie superalgebras, and mechanics.
\newblock {\em Dokl. Akad. Nauk SSSR}, 236(4):804--807, 1977.

\bibitem{Lei80}
D.~A. Le{\u\i}tes.
\newblock Introduction to the theory of supermanifolds.
\newblock {\em Uspekhi Mat. Nauk}, 35(1(211)):3--57, 255, 1980.

\bibitem{MRa09}
P.~Mathonet and F.~Radoux.
\newblock {On natural and conformally equivariant quantizations.}
\newblock {\em J. Lond. Math. Soc., II. Ser.}, 80(1):256--272, 2009.

\bibitem{Mic09}
J.-P. Michel.
\newblock {\em Quantification conform{\'e}ment {\'e}quivariante des fibr{\'e}s
  supercotangents}.
\newblock PhD thesis, Universit{\'e} Aix-Marseille II, 2009.
\newblock Electronically available as tel-00425576.

\bibitem{Mic11a}
J.-Ph. Michel.
\newblock Conformally equivariant quantization - a complete classification.
\newblock {\em SIGMA}, 8:Paper 022, 2012.
\newblock arXiv:1102.4065.

\bibitem{MPU08}
I.~M. Musson, G.~Pinczon, and R.~Ushirobira.
\newblock Hochschild cohomology and deformations of clifford-weyl algebras.
\newblock {\em Sigma}, 5:Paper 028, 2009.

\bibitem{NTr02}
P.~Nurowski and A.~Trautman.
\newblock Robinson manifolds as the {L}orentzian analogs of {H}ermite
  manifolds.
\newblock {\em Differential Geom. Appl.}, 17(2-3):175--195, 2002.
\newblock 8th International Conference on Differential Geometry and its
  Applications (Opava, 2001).

\bibitem{ORe03}
V.~Yu. Ovsienko and P.~Redou.
\newblock Generalized transvectants-{R}ankin-{C}ohen brackets.
\newblock {\em Lett. Math. Phys.}, 63(1):19--28, 2003.

\bibitem{PWi08}
M.~Palese and E.~Winterroth.
\newblock Noether identities in {E}instein-{D}irac theory and the {L}ie
  derivative of spinor fields.
\newblock In {\em Differential geometry and its applications}, pages 643--653.
  World Sci. Publ., Hackensack, NJ, 2008.

\bibitem{Pan08}
S.~M. Paneitz.
\newblock A quartic conformally covariant differential operator for arbitrary
  pseudo-{R}iemannian manifolds (summary).
\newblock {\em SIGMA}, 4:Paper 036, 2008.

\bibitem{Pap51}
A.~Papapetrou.
\newblock Spinning test-particles in general relativity. {I}.
\newblock {\em Proc. Roy. Soc. London. Ser. A.}, 209:248--258, 1951.

\bibitem{Rad09}
F.~Radoux.
\newblock An explicit formula for the natural and conformally invariant
  quantization.
\newblock {\em Lett. Math. Phys.}, 89(3):249--263, 2009.

\bibitem{Rav80}
F.~Ravndal.
\newblock Supersymmetric {D}irac particles in external fields.
\newblock {\em Phys. Rev. D (3)}, 21(10):2823--2832, 1980.

\bibitem{Rot90}
M.~Rothstein.
\newblock The structure of supersymplectic supermanifolds.
\newblock In {\em Differential geometric methods in theoretical physics
  ({R}apallo, 1990)}, volume 375 of {\em Lecture Notes in Phys.}, pages
  331--343. Springer, Berlin, 1991.

\bibitem{Roy02}
D.~Roytenberg.
\newblock On the structure of graded symplectic supermanifolds and {C}ourant
  algebroids.
\newblock In {\em Quantization, {P}oisson brackets and beyond ({M}anchester,
  2001)}, volume 315 of {\em Contemp. Math.}, pages 169--185. Amer. Math. Soc.,
  Providence, RI, 2002.

\bibitem{Sil09}
J.~Silhan.
\newblock Conformally invariant quantization - towards complete classification.
\newblock 2009.

\bibitem{Sou70}
J.-M. Souriau.
\newblock {\em Structure des syst{\`e}mes dynamiques}.
\newblock Ma{\^i}trise de math{\'e}matiques. Dunod, Paris, 1970 ($\copyright$
  1969).

\bibitem{HSp01}
G.~A.~J. Sparling and J.~E. Holland.
\newblock Conformally invariant powers of the ambient dirac operator.
\newblock {\em arXiv:math/0112033}, 2001.

\bibitem{Tra08}
A.~{Trautman}.
\newblock Connections and the dirac operator on spinor bundles.
\newblock {\em Journal of Geometry and Physics}, 58:238--252, February 2008.

\bibitem{Tuy92}
G.~M. Tuynman.
\newblock Geometric quantization of the {BRST} charge.
\newblock {\em Comm. Math. Phys.}, 150(2):237--265, 1992.

\bibitem{Tuy04}
G.~M. Tuynman.
\newblock {\em Supermanifolds and supergroups}, volume 570 of {\em Mathematics
  and its Applications}.
\newblock Kluwer Academic Publishers, Dordrecht, 2004.
\newblock Basic theory.

\bibitem{Vor90}
F.~F. Voronov.
\newblock Quantization on supermanifolds and an analytic proof of the
  {A}tiyah-{S}inger index theorem.
\newblock In {\em Current problems in mathematics. {N}ewest results, {V}ol.\ 38
  ({R}ussian)}, Itogi Nauki i Tekhniki, pages 3--118, 186. Akad. Nauk SSSR
  Vsesoyuz. Inst. Nauchn. i Tekhn. Inform., Moscow, 1990.
\newblock Translated in J. Soviet Math. {{\bf{6}}4} (1993), no. 4, 993--1069.

\bibitem{Wey97}
H.~Weyl.
\newblock {\em The classical groups}.
\newblock Princeton Landmarks in Mathematics. Princeton University Press,
  Princeton, NJ, 1997.
\newblock Their invariants and representations, Fifteenth printing, Princeton
  Paperbacks.

\end{thebibliography}
 
\end{document}